\input pictex.tex   %% Dovrebbe essere in path (altrimenti vedi in Doc)

% $Id: dcpic.sty,v 1.24 2002/11/25 13:51:57 pedro Exp $
%% DC-PiCTeX
%% Realizado por Pedro Quaresma de Almeida, Coimbra 
%% 11/1990 (vers{\~a}o 1.0); 10/1991 (vers{\~a}o 1.1);
%%  9/1993 (vers{\~a}o 1.2);  3/1995 (vers{\~a}o 1.3);
%%  7/1996 (vers{\~a}o 2.1);
%%  5/2001 (vers{\~a}o 3.0); 11/2001 (vers{\~a}o 3.1);
%%  1/2002 (vers{\~a}o 3.2)
%%  5/2002 (versão 4.0)
\immediate\write10{Package DCpic 2002/05/16 v4.0}

\catcode`!=11 %  ***** THIS MUST NEVER BE OMITTED (Ver PiCTeX)

\newcount\aux%
\newcount\auxa%
\newcount\auxb%
\newcount\m%
\newcount\n%
\newcount\x%
\newcount\y%
\newcount\xl%
\newcount\yl%
\newcount\d%
\newcount\dnm%
\newcount\xa%
\newcount\xb%
\newcount\xmed%
\newcount\xc%
\newcount\xd%
\newcount\ya%
\newcount\yb%
\newcount\ymed%
\newcount\yc%
\newcount\yd
%% "variáveis globais"
\newcount\expansao%
\newcount\tipografo%       versão 4.0
\newcount\distanciaobjmor% versão 4.0
\newcount\tipoarco%        versão 4.0
%\newif\ifarredondada%        versão 4.0 (valor inicial "false")
\newif\ifpara%
%% version 3.2
\newbox\caixa%
\newbox\caixaaux%
\newif\ifnvazia%
\newif\ifvazia%
\newif\ifcompara%
\newif\ifdiferentes%
\newcount\xaux%
\newcount\yaux%
\newcount\guardaauxa%
\newcount\alt%
\newcount\larg%
\newcount\prof%
%% para os ajustes
\newcount\auxqx
\newcount\auxqy
\newif\ifajusta%
\newif\ifajustadist
\def\objPartida{}%
\def\objChegada{}%
\def\objNulo{}%

%% 
%% Stack specification
%%

%%
%% Emtpy stack
%%
\def\!vazia{:}

%%
%% Is Empty? : Stack -> Bool
%%
%% nvazia - True if Not Empy
%% vazia  - True if Empty
\def\!pilhanvazia#1{\let\arg=#1%
\if:\arg\ \nvaziafalse\vaziatrue \else \nvaziatrue\vaziafalse\fi}

%%
%% Push : Elems x Stack -> Stack
%%
\def\!coloca#1#2{\edef\pilha{#1.#2}}

%%
%% Top : Stack -> Elems
%%
%% the empty stack is not taken care
%% the element is "kept" ("guardado") 
\def\!guarda(#1)(#2,#3)(#4,#5,#6){\def\id{#1}%
\xaux=#2%
\yaux=#3%
\alt=#4%
\larg=#5%
\prof=#6%
}

\def\!topaux#1.#2:{\!guarda#1}
\def\!topo#1{\expandafter\!topaux#1}

%%
%% Pop : Stack -> Stack
%%
%% the empty stack is not taken care
\def\!popaux#1.#2:{\def\pilha{#2:}}
\def\!retira#1{\expandafter\!popaux#1}

%%
%% Compares words : Word x Word -> Bool
%%
%% compara - True if equal
%% diferentes - True if not equal
\def\!comparaaux#1#2{\let\argA=#1\let\argB=#2%
\ifx\argA\argB\comparatrue\diferentesfalse\else\comparafalse\diferentestrue\fi}

\def\!compara#1#2{\!comparaaux{#1}{#2}}

%%Comando Interno
%% Valor absoluto (absolute value)
%% \absoluto{n}{absn}
%% entrada
%%  n - natural
%% sa{\'\i}da
%%  absn - o valor absoluto de n
\def\!absoluto#1#2{\n=#1%
  \ifnum \n > 0
    #2=\n
  \else
    \multiply \n by -1
    #2=\n
  \fi}

%% Name definitions for edge types and directions

\def\solidline{2}

%% Name definitions for edge label placement

\def\atleft{1}
%% Tip direction for curved edges

%% Type of graph
\def\commdiag{0}

%% Posicionamento da etiquetas nos grafos

%%Comando Interno
%% Ajusta a dist{\^a}ncia entre as setas e os objectos em fun{\c c}{\~a}o das
%% dimens{\~o}es destes {\'u}ltimos
%% \ajusta{x}{xl}{y}{yl}{d}{Objecto}
%% entrada
%%  (x,y) e (xl,yl), coordenadas dos pontos de {\'\i}nicio e fim da seta
%%  d, dist{\^a}ncia especificada pelo utilizador ou 10 (valor por
%%  omiss{\~a}o), Objecto d{\'a}-nos a refer{\^e}ncia do objecto ao qual se est{\'a} a
%%  efectuar o ajuste.
%% sa{\'\i}da
%%  d, dist{\^a}ncia alterada.
%% 
%% A dist{\^a}ncia alterada {\'e} o maior valor entre 10 e as dimens{\~o}es
%% apropriadas da caixa que cont{\^e}m o objecto. 
%% Se o utilizador especificar um valor essa especifica{\c c}{\~a}o
%% n{\~a}o {\'e} alterada.
%%
%% Se a seta {\'e} horizontal usa-se o valor da largura
%% Se a seta {\'e} vertical usa-se:
%%  o valor da altura se a seta est{\'a} no 1o ou 2o quadrante
%%  o valor da profundidade se a seta est{\'a} no 3o ou 4o quadrante
%% Se a seta {\'e} {\'o}bliqua vai-se escolher o valor conforme:
%%  de 315 a  45 graus usa-se a largura
%%  de  45 a 135 graus usa-se a altura
%%  de 135 a 225 graus usa-se a largura
%%  de 225 a 315 graus usa-se a profundidade
\def\!ajusta#1#2#3#4#5#6{\aux=#5%
  \let\auxobj=#6%
  \ifcase \tipografo    % diagramas comutativos
    \ifnum\number\aux=10 
      \ajustadisttrue % se o valor é o valor por omissão ajusta
    \else
      \ajustadistfalse  % caso contrário não ajusta
    \fi
  \else  % grafos (dirigidos, não dirigidos, com molduras)
   \ajustadistfalse
%  \or  % grafos não dirigidos
%   \ajustadistfalse
%  \else % grafos dirigidos com molduras circulares nos objectos
%    \ifnum\number\aux=8 
%      \ajustadisttrue  % se o valor é o valor por omissão ajusta
%    \else
%      \ajustadistfalse % caso contrário não ajusta
%    \fi
  \fi
  \ifajustadist
%  \tiny Vou ajustar %%%%%%%%%%%%%%%%%%%%%%%%%%%%%%
%  \ifnum\number\aux=10% verificar se s{\~a}o os valores por omiss{\~a}o
   \let\pilhaaux=\pilha%
   \loop%
     \!topo{\pilha}%
     \!retira{\pilha}%
     \!compara{\id}{\auxobj}%
     \ifcompara\nvaziafalse \else\!pilhanvazia\pilha \fi%
     \ifnvazia%
   \repeat%
%% rep{\~o}e os valores na pilha
   \let\pilha=\pilhaaux%
   \ifvazia%
    \ifdiferentes%
%%
%% N{\~a}o {\'e} poss{\'\i}vel efectuar o ajuste dado o utilizador n{\~a}o ter
%% especificado uma etiqueta para o objecto em quest{\~a}o. {\'E} dado o
%% valor de 10, igual ao valor por omiss{\~a}o.
%%
     \larg=1310720% n{\~a}o faz o ajuste
     \prof=655360%
     \alt=655360%
    \fi%
   \fi%
   \divide\larg by 131072
   \divide\prof by 65536
   \divide\alt by 65536
   \ifnum\number\y=\number\yl
%% Caso 1 -- seta horizontal
%%
%% divide-se por 131072 para se obter metade da largura da caixa em
%% pontos (pt), isto dado que o texto est{\'a} centrado na caixa. Soma-se
%% mais tr{\^e}s, que constitue um ajuste imp{\'\i}rico.
    \advance\larg by 3
    \ifnum\number\larg>\aux
     #5=\larg
    \fi
   \else
    \ifnum\number\x=\number\xl
     \ifnum\number\yl>\number\y
%% Caso 2.1 -- seta vertical de cima para baixa
%%
      \ifnum\number\alt>\aux
       #5=\alt
      \fi
     \else
%% Caso 2.2 -- seta vertical de baixo para cima
%%
%% divide-se por 65536 para se obter a altura da caixa em pt. O ajuste
%% de 5 foi obtido imp{\'\i}ricamente
      \advance\prof by 5
      \ifnum\number\prof>\aux
       #5=\prof
      \fi
     \fi
    \else
%% Caso 3 -- seta obl{\'\i}qua 
%% Caso 3.1 de 315o a  45o; |x-xl|>|y-yl| e
%% Caso 3.3 de 135o a 225o; |x-xl|>|y-yl|; Largura
     \auxqx=\x
     \advance\auxqx by -\xl
     \!absoluto{\auxqx}{\auxqx}%
     \auxqy=\y
     \advance\auxqy by -\yl
     \!absoluto{\auxqy}{\auxqy}%
     \ifnum\auxqx>\auxqy
      \ifnum\larg<10
       \larg=10
      \fi
      \advance\larg by 3
      #5=\larg
     \else
%% Caso 3.2 de  45o a 135o; |x-xl|<|y-yl| e y>0; Largura
      \ifnum\yl>\y
       \ifnum\larg<10
        \larg=10
       \fi
      \advance\alt by 6
       #5=\alt
      \else
%% Caso 3.4 de 225o a 315o; |x-xl|<|y-yl| e y<0; Profundidade
      \advance\prof by 11
       #5=\prof
      \fi
     \fi
    \fi
   \fi
\fi} % o ramo "else" {\'e} omisso

%%Comando Interno
%% C{\'a}lculo da Raiz Quadrada
%% raiz{n}{m}
%% entrada
%%   n - natural
%% sa{\'\i}da
%%   n - natural
%%   m - maior natural contido na raiz quadrada de n
\def\!raiz#1#2{\n=#1%
  \m=1%
  \loop
    \aux=\m%
    \advance \aux by 1%
    \multiply \aux by \aux%
    \ifnum \aux < \n%
      \advance \m by 1%
      \paratrue%
    \else\ifnum \aux=\n%
      \advance \m by 1%
      \paratrue%
       \else\parafalse%
       \fi
    \fi
  \ifpara%
  \repeat
#2=\m}

%%Comando Interno
%% Calcula os pontos de 
%%       come{\c c}o da "seta"
%%       fim da "seta"
%%   coloca{\c c}{\~a}o do s{\'\i}mbolo
%% 
%% ucoord{x1}{x2}{x3}{x4}{x5}{x6}{+|- 1}
%% entrada
%%   x1,x2,x3,x4,x5
%% sa{\'\i}da
%%   x6
%%  
%%              x2 - x1
%%  x6 = x3 +|- ------- x4
%%                 x5
\def\!ucoord#1#2#3#4#5#6#7{\aux=#2%
  \advance \aux by -#1%
  \multiply \aux by #4%
  \divide \aux by #5%
  \ifnum #7 = -1 \multiply \aux by -1 \fi%
  \advance \aux by #3%
#6=\aux}

%%Comando Interno 
%% C{\'a}lculo do Quadrado da Dist{\^a}ncia Euclidiana entre dois pontos 
%% quadrado{n}{m}{l}
%% entrada
%%   n - natural
%%   m - natural
%% sa{\'\i}da
%%   l = (n-m)*(n-m)
\def\!quadrado#1#2#3{\aux=#1%
  \advance \aux by -#2%
  \multiply \aux by \aux%
#3=\aux}

%%Comando Interno
%% C{\'a}lculo auxiliar para determinar a dist{\^a}ncia entre o nome do
%% morfismo e a seta.
%% entrada
%%     (x,y), (x',y') e o nome do morfismo
%% sa{\'\i}da
%%     dnm - dist{\^a}ncia do nome ao morfismo respectivo devidamente
%%     compensada pelo tamanho do objecto
%% Observa{\c c}{\~o}es
%%     A compensa{\c c}{\~a}o s{\'o} est{\'a} a ser feita para setas
%%     horizontais e verticais. As obl{\'\i}quas s{\~a}o tratadas de
%%     outra forma.
%% algoritmo
%%  caixa0 <- nome do morfismo
%%  se x-xl = 0 entao                   {recta vertical}
%%     aux <- largura da caixa0
%%     dnm <- convers{\~a}o-sp-pt(aux)/2+3
%%  sen{\~a}o                               {recta n{\~a}o vertical}
%%     se y-yl = 0 entao                {recta horizontal}
%%        aux <- altura+profundidade da caixa0
%%        dnm <- convers{\~a}o-sp-pt(aux)/2+3
%%     sen{\~a}o                            {recta obl{\'\i}qua}
%%        dnm <- 3
%%     fimse
%%  fimse
%% fimalgoritmo
\def\!distnomemor#1#2#3#4#5#6{\setbox0=\hbox{#5}%
  \aux=#1
  \advance \aux by -#3
  \ifnum \aux=0
     \aux=\wd0 \divide \aux by 131072
     \advance \aux by 3
     #6=\aux
  \else
     \aux=#2
     \advance \aux by -#4
     \ifnum \aux=0
        \aux=\ht0 \advance \aux by \dp0 \divide \aux by 131072
        \advance \aux by 3
        #6=\aux%
     \else
     #6=3
     \fi
   \fi
}

%%
%% O ambiente "begindc...enddc"
%%
\def\begindc#1{\!ifnextchar[{\!begindc{#1}}{\!begindc{#1}[30]}}
\def\!begindc#1[#2]{\beginpicture 
  \let\pilha=\!vazia
  \setcoordinatesystem units <1pt,1pt>
  \expansao=#2
  \ifcase #1
    \distanciaobjmor=10
    \tipoarco=0         % seta
    \tipografo=0        % diagrama comutativo
  \or
    \distanciaobjmor=2
    \tipoarco=0         % seta 
    \tipografo=1        % grafo orientado
  \or
    \distanciaobjmor=1
    \tipoarco=2         % linha
    \tipografo=2        % grafo não orientado
  \or
    \distanciaobjmor=8
    \tipoarco=0         % seta 
    \tipografo=3        % grafo orientado
%    \arredondadotrue    % objectos com molduras circulares
  \or
    \distanciaobjmor=8
    \tipoarco=2         % linha
    \tipografo=4        % grafo não orientado
%    \arredondadotrue    % objectos com molduras circulares
  \fi}

\def\enddc{\endpicture}

%%
%% Comando para construir a "seta" entre dois objectos
%%
%% Os pontos definidores da seta e da etiqueta respectiva s{\~a}o:
%% 
%%                (xd,yd)
%%                   o
%%                   |
%%  o------o---------o---------o------o
%%(x,y) (xa,ya)   (xm,ym)   (xb,yb)(xl,yl)
%%
\def\mor{%
  \!ifnextchar({\!morxy}{\!morObjA}}
\def\!morxy(#1,#2){%
  \!ifnextchar({\!morxyl{#1}{#2}}{\!morObjB{#1}{#2}}}
\def\!morxyl#1#2(#3,#4){%
  \!ifnextchar[{\!mora{#1}{#2}{#3}{#4}}{\!mora{#1}{#2}{#3}{#4}[\number\distanciaobjmor,\number\distanciaobjmor]}}%
\def\!morObjA#1{%
 \let\pilhaaux=\pilha%
 \def\objPartida{#1}%
 \loop%
    \!topo\pilha%
    \!retira\pilha%
    \!compara{\id}{\objPartida}%
    \ifcompara \nvaziafalse \else \!pilhanvazia\pilha \fi%
   \ifnvazia%
 \repeat%
 \ifvazia%
  \ifdiferentes%
%%
%% Mensagem de erro e atribui{\c c}{\~a}o de valores fict{\'\i}cios aos 
%% argumentos dos comandos que se seguem.
%%
   Error: Incorrect label specification%
   \xaux=1%
   \yaux=1%
  \fi%
 \fi% 
 \let\pilha=\pilhaaux%
 \!ifnextchar({\!morxyl{\number\xaux}{\number\yaux}}{\!morObjB{\number\xaux}{\number\yaux}}}
\def\!morObjB#1#2#3{%
  \x=#1
  \y=#2
 \def\objChegada{#3}%
 \let\pilhaaux=\pilha%
 \loop
    \!topo\pilha %
    \!retira\pilha%
    \!compara{\id}{\objChegada}%
    \ifcompara \nvaziafalse \else \!pilhanvazia\pilha \fi
   \ifnvazia
 \repeat
 \ifvazia
  \ifdiferentes%
%%
%% Mensagem de erro e atribui{\c c}{\~a}o de valores fict{\'\i}cios aos 
%% argumentos dos comandos que se seguem.
%%
   Error: Incorrect label specification
   \xaux=\x%
   \advance\xaux by \x%
   \yaux=\y%
   \advance\yaux by \y%
  \fi
 \fi
 \let\pilha=\pilhaaux
 \!ifnextchar[{\!mora{\number\x}{\number\y}{\number\xaux}{\number\yaux}}{\!mora{\number\x}{\number\y}{\number\xaux}{\number\yaux}[\number\distanciaobjmor,\number\distanciaobjmor]}}
\def\!mora#1#2#3#4[#5,#6]#7{%
  \!ifnextchar[{\!morb{#1}{#2}{#3}{#4}{#5}{#6}{#7}}{\!morb{#1}{#2}{#3}{#4}{#5}{#6}{#7}[1,\number\tipoarco] }}
\def\!morb#1#2#3#4#5#6#7[#8,#9]{\x=#1%
  \y=#2%
  \xl=#3%
  \yl=#4%
  \multiply \x by \expansao%
  \multiply \y by \expansao%
  \multiply \xl by \expansao%
  \multiply \yl by \expansao%
%%
%% calcular a dist{\^a}ncia Euclidiana entre dois pontos
%% d = \sqrt((x-xl)^2+(y-yl)^2)
%%
  \!quadrado{\number\x}{\number\xl}{\auxa}%
  \!quadrado{\number\y}{\number\yl}{\auxb}%
  \d=\auxa%
  \advance \d by \auxb%
  \!raiz{\d}{\d}%
%%
%% o ponto (xa,ya) est{\'a} {\`a} dist{\^a}ncia #5 (valor por omiss{\~a}o 10) do ponto
%% (x,y)
%%
%% como existem dois pontos em considera{\c c}{\~a}o, o ponto de partida e o
%% ponto de chegada, vai sei necess{\'a}rio recuperar de novo os seus
%% valores por pesquisa na pilha
  \auxa=#5
  \!compara{\objNulo}{\objPartida}%
  \ifdiferentes% S{\'o} vai fazer o ajuste quando {\'e} necess{\'a}rio
   \!ajusta{\x}{\xl}{\y}{\yl}{\auxa}{\objPartida}%
   \ajustatrue
   \def\objPartida{}% re-inicializar o valor do Objecto de Partida
  \fi
%% vai guardar o valor de auxa (ap{\'o}s ajuste) para ser usado no caso
%% dos morfismos de injec{\c c}{\~a}o.
  \guardaauxa=\auxa
  \!ucoord{\number\x}{\number\xl}{\number\x}{\auxa}{\number\d}{\xa}{1}%
  \!ucoord{\number\y}{\number\yl}{\number\y}{\auxa}{\number\d}{\ya}{1}%
%% auxa vai ter o valor da dist{\^a}ncia entre os objectos menos a
%% dist{\^a}ncia da seta ao objecto (10 por omiss{\~a}o)
  \auxa=\d%
%%
%% o ponto (xb,yb) est{\'a} {\`a} dist{\^a}ncia #6 (valor por omiss{\~a}o 10) do ponto
%% (xl,yl)
%%
  \auxb=#6
  \!compara{\objNulo}{\objChegada}%
  \ifdiferentes% S{\'o} vai fazer o ajuste quando {\'e} necess{\'a}rio
%   Vou ajustar
   \!ajusta{\x}{\xl}{\y}{\yl}{\auxb}{\objChegada}%
   \def\objChegada{}% re-inicializar o valor do Objecto de Chegada
  \fi
  \advance \auxa by -\auxb%
  \!ucoord{\number\x}{\number\xl}{\number\x}{\number\auxa}{\number\d}{\xb}{1}%
  \!ucoord{\number\y}{\number\yl}{\number\y}{\number\auxa}{\number\d}{\yb}{1}%
  \xmed=\xa%
  \advance \xmed by \xb%
  \divide \xmed by 2
  \ymed=\ya%
  \advance \ymed by \yb%
  \divide \ymed by 2
  \!distnomemor{\number\x}{\number\y}{\number\xl}{\number\yl}{#7}{\dnm}%
  \!ucoord{\number\y}{\number\yl}{\number\xmed}{\number\dnm}{\number\d}{\xc}{-#8}% 
  \!ucoord{\number\x}{\number\xl}{\number\ymed}{\number\dnm}{\number\d}{\yc}{#8}%
\ifcase #9  % seta s{\'o}lida
  \arrow <4pt> [.2,1.1] from {\xa} {\ya} to {\xb} {\yb}
\or  % seta a tracejado
  \setdashes
  \arrow <4pt> [.2,1.1] from {\xa} {\ya} to {\xb} {\yb}
  \setsolid
\or  % linha s{\'o}lida
  \setlinear
  \plot {\xa} {\ya}  {\xb} {\yb} /
\or  % seta de injec{\c c}{\~a}o
%% C{\'a}lculos auxiliares
%%
%% 3 valor para o raio do "rabo" da "seta"
%%
%% repor o valor de auxa
  \auxa=\guardaauxa
%% dar a compensa{\c c}{\~a}o para o "rabo"
  \advance \auxa by 3%
%%
%% IMPORTANTE os valores de xa e ya v{\~a}o ser alterados
%%
 \!ucoord{\number\x}{\number\xl}{\number\x}{\number\auxa}{\number\d}{\xa}{1}%
 \!ucoord{\number\y}{\number\yl}{\number\y}{\number\auxa}{\number\d}{\ya}{1}%
 \!ucoord{\number\y}{\number\yl}{\number\xa}{3}{\number\d}{\xd}{-1}%
 \!ucoord{\number\x}{\number\xl}{\number\ya}{3}{\number\d}{\yd}{1}%
%% Constru{\c c}{\~a}o da "seta"
  \arrow <4pt> [.2,1.1] from {\xa} {\ya} to {\xb} {\yb}
%% e do seu "rabo"
  \circulararc -180 degrees from {\xa} {\ya} center at {\xd} {\yd}
\or  % seta de aplica{\c c}{\~a}o ("|-->")
  \auxa=3% valor para o meio-segmento do "rabo" da "seta"
%% c{\'a}lculo dos pontos (xmed,ymed) e (xd,yd) para o segmento de recta que
%% define o "rabo" da seta
 \!ucoord{\number\y}{\number\yl}{\number\xa}{\number\auxa}{\number\d}{\xmed}{-1}%
 \!ucoord{\number\x}{\number\xl}{\number\ya}{\number\auxa}{\number\d}{\ymed}{1}%
 \!ucoord{\number\y}{\number\yl}{\number\xa}{\number\auxa}{\number\d}{\xd}{1}%
 \!ucoord{\number\x}{\number\xl}{\number\ya}{\number\auxa}{\number\d}{\yd}{-1}%
%% Constru{\c c}{\~a}o da "seta"
  \arrow <4pt> [.2,1.1] from {\xa} {\ya} to {\xb} {\yb}
%% e do seu "rabo"
  \setlinear
  \plot {\xmed} {\ymed}  {\xd} {\yd} /
\fi
%% Coloca{\c c}{\~a}o do nome do morfismo.
%% Se os morfismos s{\~a}o horizontais ou verticais constro{\'\i} uma caixa
%% centrada no ponto pr{\'e}viamente calculado. Se as setas s{\~a}o
%% obl{\'\i}quas coloca a caixa de forma a n{\~a}o colidir com o morfismo 
%% tendo em aten{\c c}{\~a}o o quadrante assim como a posi{\c c}{\~a}o
%% relativa do morfismo e do respectivo nome.
\auxa=\xl
\advance \auxa by -\x%
\ifnum \auxa=0 
  \put {#7} at {\xc} {\yc}
\else
  \auxb=\yl
  \advance \auxb by -\y%
  \ifnum \auxb=0 \put {#7} at {\xc} {\yc}
  \else 
    \ifnum \auxa > 0 
      \ifnum \auxb > 0
        \ifnum #8=1
          \put {#7} [rb] at {\xc} {\yc}
        \else 
          \put {#7} [lt] at {\xc} {\yc}
        \fi
      \else
        \ifnum #8=1
          \put {#7} [lb] at {\xc} {\yc}
        \else 
          \put {#7} [rt] at {\xc} {\yc}
        \fi
      \fi
    \else
      \ifnum \auxb > 0 
        \ifnum #8=1
          \put {#7} [rt] at {\xc} {\yc}
        \else 
          \put {#7} [lb] at {\xc} {\yc}
        \fi
      \else
        \ifnum #8=1
          \put {#7} [lt] at {\xc} {\yc}
        \else 
          \put {#7} [rb] at {\xc} {\yc}
        \fi
      \fi
    \fi
  \fi
\fi
}

%%
%% Comando para construir a "seta" curvilinea entre dois objectos
%%
%% \cmor(<lista de pontos (n. impar)>){<etiqueta>}
%%
%% Em primeiro lugar {\'e} necess{\'a}rio modificar o comando plot de forma a
%% que a sintaxe de utiliza{\c c}{\~a}o do novo comando seja coerente com a
%% sintaxe dos restantes comandos
%%
\def\modifplot(#1{\!modifqcurve #1}
\def\!modifqcurve(#1,#2){\x=#1%
  \y=#2%
  \multiply \x by \expansao%
  \multiply \y by \expansao%
  \!start (\x,\y)
  \!modifQjoin}
\def\!modifQjoin(#1,#2)(#3,#4){\x=#1%
  \y=#2%
  \xl=#3%
  \yl=#4%
  \multiply \x by \expansao%
  \multiply \y by \expansao%
  \multiply \xl by \expansao%
  \multiply \yl by \expansao%
  \!qjoin (\x,\y) (\xl,\yl)             % \!qjoin  is defined in QUADRATIC
  \!ifnextchar){\!fim}{\!modifQjoin}}
\def\!fim){\ignorespaces}

%%
%% O comando para desenhar a seta vai receber a lista de pontos da qual
%% retira o {\'u}ltimo par de pontos, dependente da escolha dada pelo
%% utilizador a seta vai ser desenhada para cima, para baixo, para a
%% direita ou para a esquerda
%%
\def\setaxy(#1{\!pontosxy #1}
\def\!pontosxy(#1,#2){%
  \!maispontosxy}
\def\!maispontosxy(#1,#2)(#3,#4){%
  \!ifnextchar){\!fimxy#3,#4}{\!maispontosxy}}
\def\!fimxy#1,#2){\x=#1%
  \y=#2
  \multiply \x by \expansao
  \multiply \y by \expansao
  \xl=\x%
  \yl=\y%
  \aux=1%
  \multiply \aux by \auxa%
  \advance\xl by \aux%
  \aux=1%
  \multiply \aux by \auxb%
  \advance\yl by \aux%
  \arrow <4pt> [.2,1.1] from {\x} {\y} to {\xl} {\yl}}

%%
%% Temos agora a defini{\c c}{\~a}o do comando "cmor"
%%
\def\cmor#1 #2(#3,#4)#5{%
  \!ifnextchar[{\!cmora{#1}{#2}{#3}{#4}{#5}}{\!cmora{#1}{#2}{#3}{#4}{#5}[0] }}
\def\!cmora#1#2#3#4#5[#6]{%
  \ifcase #2% para cima "\pup" (pointing up)
      \auxa=0% x mant{\^e}m-se
      \auxb=1% o y "sobe" 
    \or% para baixo "\pdown" (pointing down)
      \auxa=0% x mant{\^e}m-se
      \auxb=-1% o y "desce" 
    \or% para a direita "\pright" (pointing right)
      \auxa=1% o x move-se para a direita
      \auxb=0% o y mant{\^e}m-se
    \or% para a esquerda "\pleft" (pointing left)
      \auxa=-1% o x move-se para a esquerda
      \auxb=0% o y mant{\^e}m-se
    \fi  % constru{\c c}{\~a}o do arco
  \ifcase #6  % arco (com seta) s{\'o}lido
    \modifplot#1% Desenhar o arco
    % constru{\c c}{\~a}o da seta
    \setaxy#1
  \or  % arco (com seta) a tracejado
    \setdashes
    \modifplot#1% Desenhar o arco
    \setaxy#1
    \setsolid
  \or  % arco sem seta
    \modifplot#1% Desenhar o arco
  \fi  % seta de injec{\c c}{\~a}o
%% coloca{\c c}{\~a}o da etiqueta do morfismo
  \x=#3%  
  \y=#4%
  \multiply \x by \expansao%
  \multiply \y by \expansao%
  \put {#5} at {\x} {\y}}

%%
%% Comando para construir os Objectos
%%  \obj(x,y){<texto>}[<etiqueta>]
%% 
\def\obj(#1,#2){%
  \!ifnextchar[{\!obja{#1}{#2}}{\!obja{#1}{#2}[Nulo]}}
\def\!obja#1#2[#3]#4{%
  \!ifnextchar[{\!objb{#1}{#2}{#3}{#4}}{\!objb{#1}{#2}{#3}{#4}[1]}}
\def\!objb#1#2#3#4[#5]{%
  \x=#1%
  \y=#2%
  \def\!pinta{\normalsize$\bullet$}% para definir o tamanho normal das pintas
  \def\!nulo{Nulo}%
  \def\!arg{#3}%
  \!compara{\!arg}{\!nulo}%
  \ifcompara\def\!arg{#4}\fi%
  \multiply \x by \expansao%
  \multiply \y by \expansao%
  \setbox\caixa=\hbox{#4}%
  \!coloca{(\!arg)(#1,#2)(\number\ht\caixa,\number\wd\caixa,\number\dp\caixa)}{\pilha}%
  \auxa=\wd\caixa \divide \auxa by 131072 
  \advance \auxa by 5
  \auxb=\ht\caixa
  \advance \auxb by \number\dp\caixa
  \divide \auxb by 131072 
  \advance \auxb by 5
%(\number\auxa,
%\number\auxb)
%  \aux=\ht\caixa \divide \auxa by 131072 
% \advance \auxa by 5 
%  \auxb=\dp\caixa \divide \auxb by 131072 
%  \advance \auxb by 8
  \ifcase \tipografo    % diagramas comutativos
    \put{#4} at {\x} {\y}
  \or                   % grafos dirigidos
    \ifcase #5 % c=0
      \put{#4} at {\x} {\y}
    \or        % n=1
      \put{\!pinta} at {\x} {\y}
      \advance \y by \number\auxb  % height+depth+5
      \put{#4} at {\x} {\y}
    \or        % ne=2
      \put{\!pinta} at {\x} {\y}
      \advance \auxa by -2  % para fazer o ajuste (imperfeito)
      \advance \auxb by -2  % ao raio da circunferência de centro (x,y)
      \advance \x by \number\auxa  % width+5
      \advance \y by \number\auxb  % height+depth+5
      \put{#4} at {\x} {\y}   
    \or        % e=3
      \put{\!pinta} at {\x} {\y}
      \advance \x by \number\auxa  % width+5
      \put{#4} at {\x} {\y}   
    \or        % se=4
      \put{\!pinta} at {\x} {\y}
      \advance \auxa by -2  % para fazer o ajuste (imperfeito)
      \advance \auxb by -2  % ao raio da circunferência de centro (x,y)
      \advance \x by \number\auxa  % width+5
      \advance \y by -\number\auxb  % height+depth+5
      \put{#4} at {\x} {\y}   
    \or        % s=5
      \put{\!pinta} at {\x} {\y}
      \advance \y by -\number\auxb  % height+depth+5
      \put{#4} at {\x} {\y}   
    \or        % sw=6
      \put{\!pinta} at {\x} {\y}
      \advance \auxa by -2  % para fazer o ajuste (imperfeito)
      \advance \auxb by -2  % ao raio da circunferência de centro (x,y)
      \advance \x by -\number\auxa  % width+5
      \advance \y by -\number\auxb  % height+depth+5
      \put{#4} at {\x} {\y}   
    \or        % w=7
      \put{\!pinta} at {\x} {\y}
      \advance \x by -\number\auxa  % width+5
      \put{#4} at {\x} {\y}   
    \or        % nw=8
      \put{\!pinta} at {\x} {\y}
      \advance \auxa by -2  % para fazer o ajuste (imperfeito)
      \advance \auxb by -2  % ao raio da circunferência de centro (x,y)
      \advance \x by -\number\auxa  % width+5
      \advance \y by \number\auxb  % height+depth+5
      \put{#4} at {\x} {\y}   
    \fi
  \or                   % grafos não dirigidos
    \ifcase #5 % c=0
      \put{#4} at {\x} {\y}
    \or        % n=1
      \put{\!pinta} at {\x} {\y}
      \advance \y by \number\auxb  % height+depth+5
      \put{#4} at {\x} {\y}
    \or        % ne=2
      \put{\!pinta} at {\x} {\y}
      \advance \auxa by -2  % para fazer o ajuste (imperfeito)
      \advance \auxb by -2  % ao raio da circunferência de centro (x,y)
      \advance \x by \number\auxa  % width+5
      \advance \y by \number\auxb  % height+depth+5
      \put{#4} at {\x} {\y}   
    \or        % e=3
      \put{\!pinta} at {\x} {\y}
      \advance \x by \number\auxa  % width+5
      \put{#4} at {\x} {\y}   
    \or        % se=4
      \put{\!pinta} at {\x} {\y}
      \advance \auxa by -2  % ver acima
      \advance \auxb by -2
      \advance \x by \number\auxa  % width+5
      \advance \y by -\number\auxb % height+depth+5
      \put{#4} at {\x} {\y}   
    \or        % s=5
      \put{\!pinta} at {\x} {\y}
      \advance \y by -\number\auxb % height+depth+5
      \put{#4} at {\x} {\y}   
    \or        % sw=6
      \put{\!pinta} at {\x} {\y}
      \advance \auxa by -2  % ver acima
      \advance \auxb by -2
      \advance \x by -\number\auxa % width+5
      \advance \y by -\number\auxb % height+depth+5
      \put{#4} at {\x} {\y}   
    \or        % w=7
      \put{\!pinta} at {\x} {\y}
      \advance \x by -\number\auxa % width+5
      \put{#4} at {\x} {\y}   
    \or        % nw=8
      \put{\!pinta} at {\x} {\y}
      \advance \auxa by -2  % ver acima
      \advance \auxb by -2
      \advance \x by -\number\auxa % width+5
      \advance \y by \number\auxb  % height+depth+5
      \put{#4} at {\x} {\y}   
    \fi
%  \or % grafos dirigidos com molduras circulares nos objectos
%    \advance \auxa by 4
%    \put{\circle{\auxa}} [Bl] at {\x} {\y}
%    \put{#4} at {\x} {\y}
%  \or % grafos não dirigidos com molduras circulares nos objectos
   \else % grafos com molduras circulares nos objectos
     \ifnum\auxa<\auxb % determina a maior das dimensões
       \aux=\auxb
     \else
       \aux=\auxa
     \fi
% se a largura da caixa é menor do que 1em então o tamanho 
% tamanho é ajustado para esse valor mínimo
     \ifdim\wd\caixa<1em
       \dimen99 = 1em
       \aux=\dimen99 \divide \aux by 131072 
       \advance \aux by 5
     \fi
     \advance\aux by -2 %folga entre o objecto e a moldura
     \multiply\aux by 2 % 
     \ifnum\aux<30
       \put{\circle{\aux}} [Bl] at {\x} {\y}
     \else
       \multiply\auxa by 2
       \multiply\auxb by 2
       \put{\oval(\auxa,\auxb)} [Bl] at {\x} {\y}
     \fi
     \put{#4} at {\x} {\y}
   \fi   
}

\catcode`!=12 %  *****  THIS MUST NEVER BE OMITTED (Ver PiCTeX)
%\input ColorGraphicx.tex

%\input graphicx.tex
%
%    Bundle of my macros.    Version 1.2.0.beta
%    The best use is to paste all of them into the papers
%     1/8/2005
%

%
%    Fonts.    Version 1.2.0.beta
%    The best use is to paste all of them into the papers
%     1/8/2005
%
%
% History:
%	3 Agosto 2005: ChernSimons.tex
%
%%%%%%%%%%%%%%%%%%%%%%%%%%%%

%%%%%%%%%%%%%%%%%%%%%%%%%%%%%%%%%%%%%%%%%%%%%%%%%%%%
%% Font Types	%%%%%%%%%%%%%%%%%%%%%%%%%%%%%%%%%%%%%%%%%%%%
%%%%%%%%%%%%%%%%%%%%%%%%%%%%%%%%%%%%%%%%%%%%%%%%%%%%%
\def\Serif{cmr}
\def\SerifBold{cmbx}
\def\SerifItalics{cmti}
\def\SerifSlanted{cmsl}
\def\SerifBoldItalics{cmbxti}
\def\SansSerif{cmss}
\def\SansSerifBold{cmssbx}
\def\SansSerifItalics{cmssi}
\def\SansSerifSlanted{cmssi}%%
\def\Math{cmmi}
\def\Symbols{cmsy}
\def\MathBold{cmmib}
\def\MoreSymbols{cmex}
\def\Typewriter{cmtt}
\def\Gothic{eufm}
\def\Double{msbm}
\def\Relazioni{msam}

%% Font Declarations	
%\font\tenbg=cmmib10%
%\def\bg{\tenbg}%
%%%%%%%%%%%%%%%%%%%%%%%%%%%%%%%%%%%%%%%%%%%%%%%%%%%%%%
%%%	5		%%%%%%%%%%%%%%%%%%%%%%%%%%%%%%%%%%%%%%%%%%%%
%%%	%%%%%%%%%%%%%%%%%%%%%%
= 			\Serif10 			at 5pt
= 		\SerifBold10 		at 5pt
= 	\SerifItalics10 	at 5pt
=		\SerifSlanted10 	at 5pt
=	\SerifBoldItalics10	at 5pt
= 		\SansSerif10 		at 5pt
=	\SansSerifBold10	at 5pt
=	\SansSerifItalics10	at 5pt
=	\SansSerifSlanted10	at 5pt
=				\Math10				at 5pt
=			\MathBold10			at 5pt
=			\Symbols10			at 5pt
=		\MoreSymbols10		at 5pt
=		\Typewriter10		at 5pt
=			\Gothic10			at 5pt
=			\Double10			at 5pt

%%%	7		%%%%%%%%%%%%%%%%%%%%%%%%%%%%%%%%%%%%%%%%%%%
%%%	%%%%%%%%%%%%%%%%%%%%%%%
= 			\Serif10 			at 7pt
= 		\SerifBold10 		at 7pt
= 	\SerifItalics10 	at 7pt
=	\SerifSlanted10 	at 7pt
=\SerifBoldItalics10	at 7pt
= 		\SansSerif10 		at 7pt
= 	\SansSerifBold10 	at 7pt
=\SansSerifItalics10	at 7pt
=\SansSerifSlanted10	at 7pt
=			\Math10				at 7pt
=		\MathBold10			at 7pt
=			\Symbols10			at 7pt
=		\MoreSymbols10		at 7pt
=		\Typewriter10		at 7pt
=			\Gothic10			at 7pt
=			\Double10			at 7pt

%%%	8		%%%%%%%%%%%%%%%%%%%%%%%%%%%%%%%%%%%%%%%%%
%%%	%%%%%%%%%%%%%%%%%%%%%%%%%
= 			\Serif10 			at 8pt
= 		\SerifBold10 		at 8pt
= 	\SerifItalics10 	at 8pt
=	\SerifSlanted10 	at 8pt
=\SerifBoldItalics10	at 8pt
= 		\SansSerif10 		at 8pt
= 	\SansSerifBold10 	at 8pt
=\SansSerifItalics10 at 8pt
=\SansSerifSlanted10 at 8pt
=			\Math10				at 8pt
=		\MathBold10			at 8pt
=			\Symbols10			at 8pt
=		\MoreSymbols10		at 8pt
=		\Typewriter10		at 8pt
=			\Gothic10			at 8pt
=			\Double10			at 8pt

%%%	10		%%%%%%%%%%%%%%%%%%%%%%%%%%%%%%%%%%%%%
%%%	%%%%%%%%%%%%%%%%%%%%%%%%%%%%%
= 			\Serif10 			at 10pt
= 		\SerifBold10 		at 10pt
= 		\SerifItalics10 	at 10pt
=		\SerifSlanted10 	at 10pt
=	\SerifBoldItalics10	at 10pt
= 		\SansSerif10 		at 10pt
= 	\SansSerifBold10 	at 10pt
= 	\SansSerifItalics10 at 10pt
= 	\SansSerifSlanted10 at 10pt
=				\Math10				at 10pt
=			\MathBold10			at 10pt
=			\Symbols10			at 10pt
=		\MoreSymbols10		at 10pt
=		\Typewriter10		at 10pt
=			\Gothic10			at 10pt
=			\Double10			at 10pt
=			\Relazioni10			at 10pt

%%%	12		%%%%%%%%%%%%%%%%%%%%%%%%%%%%%%%%%%%%%
%%%	%%%%%%%%%%%%%%%%%%%%%%%%%%%%%
= 				\Serif10 			at 12pt
= 			\SerifBold10 		at 12pt
= 		\SerifItalics10 	at 12pt
=		\SerifSlanted10 	at 12pt
=	\SerifBoldItalics10	at 12pt
= 			\SansSerif10 		at 12pt
= 		\SansSerifBold10 	at 12pt
= 	\SansSerifItalics10 at 12pt
= 	\SansSerifSlanted10 at 12pt
=				\Math10				at 12pt
=			\MathBold10			at 12pt
=			\Symbols10			at 12pt
=		\MoreSymbols10		at 12pt
=			\Typewriter10		at 12pt
=				\Gothic10			at 12pt
=				\Double10			at 12pt

%%%	14		%%%%%%%%%%%%%%%%%%%%%%%%%%%%%%%%
= 			\Serif10 			at 14pt
= 		\SerifBold10 		at 14pt
= 	\SerifItalics10 	at 14pt
=		\SerifSlanted10 	at 14pt
=	\SerifBoldItalics10	at 14pt
= 		\SansSerif10 		at 14pt
= 	\SansSerifBold10 	at 14pt
= \SansSerifSlanted10 at 14pt
= \SansSerifItalics10 at 14pt
=				\Math10				at 14pt
=			\MathBold10			at 14pt
=			\Symbols10			at 14pt
=		\MoreSymbols10		at 14pt
=		\Typewriter10		at 14pt
=			\Gothic10			at 14pt
=			\Double10			at 14pt

%% Styles	%%%%%%%%%%%%%%%%%%%%%%%%%%%%%%%%%%%%%%%%%%%%%
%% %%%%%%%%%%%%%%%%%%%%%%%%%%%%%%%%%%%%%%%%%%%%%
\def\NormalStyle{\parindent=5pt\parskip=3pt\normalbaselineskip=14pt%
\def\nt{\tenSerif}%
\def\rm{\fam0\tenSerif}%
\textfont0=\tenSerif\scriptfont0=\sevenSerif\scriptscriptfont0=\fiveSerif%text(\tenrm)
\textfont1=\tenMath\scriptfont1=\sevenMath\scriptscriptfont1=\fiveMath%math(\tenmi)
\textfont2=\tenSymbols\scriptfont2=\sevenSymbols\scriptscriptfont2=\fiveSymbols%symbol(\tensy)
\textfont3=\tenMoreSymbols\scriptfont3=\sevenMoreSymbols\scriptscriptfont3=\fiveMoreSymbols%ex(tenex)
\textfont\itfam=\tenSerifItalics\def\it{\fam\itfam\tenSerifItalics}%
\textfont\slfam=\tenSerifSlanted\def\sl{\fam\slfam\tenSerifSlanted}%
\textfont\ttfam=\tenTypewriter\def\tt{\fam\ttfam\tenTypewriter}%
\textfont\bffam=\tenSerifBold%
\def\bf{\fam\bffam\tenSerifBold}\scriptfont\bffam=\sevenSerifBold\scriptscriptfont\bffam=\fiveSerifBold%
\def\cal{\tenSymbols}%
\def\greekbold{\tenMathBold}%
\def\gothic{\tenGothic}%
\def\Bbb{\tenDouble}%
\def\LieFont{\tenSerifItalics}%
\nt\normalbaselines\baselineskip=14pt%
}

%%%%%%%%%%%%%%%%%%%%%%%%%%%%%%%%%%%%%%%%%%%%%%%%%%%%%%%
%%%%%%%%%%%%%%%%%%%%%%
\def\TitleStyle{\parindent=0pt\parskip=0pt\normalbaselineskip=15pt%
\def\nt{\fourteenSansSerifBold}%
\def\rm{\fam0\fourteenSansSerifBold}%
\textfont0=\fourteenSansSerifBold\scriptfont0=\tenSansSerifBold\scriptscriptfont0=\eightSansSerifBold%text(\fourteenrm)
\textfont1=\fourteenMath\scriptfont1=\tenMath\scriptscriptfont1=\eightMath%math(\fourteenmi)
\textfont2=\fourteenSymbols\scriptfont2=\tenSymbols\scriptscriptfont2=\eightSymbols%symbol(\fourteensy)
\textfont3=\fourteenMoreSymbols\scriptfont3=\tenMoreSymbols\scriptscriptfont3=\eightMoreSymbols%ex(fourteenex)
\textfont\itfam=\fourteenSansSerifItalics\def\it{\fam\itfam\fourteenSansSerifItalics}%
\textfont\slfam=\fourteenSansSerifSlanted\def\sl{\fam\slfam\fourteenSerifSansSlanted}%
\textfont\ttfam=\fourteenTypewriter\def\tt{\fam\ttfam\fourteenTypewriter}%
\textfont\bffam=\fourteenSansSerif%
\def\bf{\fam\bffam\fourteenSansSerif}\scriptfont\bffam=\tenSansSerif\scriptscriptfont\bffam=\eightSansSerif%
\def\cal{\fourteenSymbols}%
\def\greekbold{\fourteenMathBold}%
\def\gothic{\fourteenGothic}%
\def\Bbb{\fourteenDouble}%
\def\LieFont{\fourteenSerifItalics}%
\nt\normalbaselines\baselineskip=15pt%
}

%%%%%%%%%%%%%%%%%%%%%%%%%%%%%%%%%%%%%%%%%%%%%%%%%%%%%%%%
%%%%%%%%%%%%%%%%%%%%%
\def\PartStyle{\parindent=0pt\parskip=0pt\normalbaselineskip=15pt%
\def\nt{\fourteenSansSerifBold}%
\def\rm{\fam0\fourteenSansSerifBold}%
\textfont0=\fourteenSansSerifBold\scriptfont0=\tenSansSerifBold\scriptscriptfont0=\eightSansSerifBold%text(\fourteenrm)
\textfont1=\fourteenMath\scriptfont1=\tenMath\scriptscriptfont1=\eightMath%math(\fourteenmi)
\textfont2=\fourteenSymbols\scriptfont2=\tenSymbols\scriptscriptfont2=\eightSymbols%symbol(\fourteensy)
\textfont3=\fourteenMoreSymbols\scriptfont3=\tenMoreSymbols\scriptscriptfont3=\eightMoreSymbols%ex(fourteenex)
\textfont\itfam=\fourteenSansSerifItalics\def\it{\fam\itfam\fourteenSansSerifItalics}%
\textfont\slfam=\fourteenSansSerifSlanted\def\sl{\fam\slfam\fourteenSerifSansSlanted}%
\textfont\ttfam=\fourteenTypewriter\def\tt{\fam\ttfam\fourteenTypewriter}%
\textfont\bffam=\fourteenSansSerif%
\def\bf{\fam\bffam\fourteenSansSerif}\scriptfont\bffam=\tenSansSerif\scriptscriptfont\bffam=\eightSansSerif%
\def\cal{\fourteenSymbols}%
\def\greekbold{\fourteenMathBold}%
\def\gothic{\fourteenGothic}%
\def\Bbb{\fourteenDouble}%
\def\LieFont{\fourteenSerifItalics}%
\nt\normalbaselines\baselineskip=15pt%
}

%%%%%%%%%%%%%%%%%%%%%%%%%%%%%%%%%%%%%%%%%%%%%%%%%%%%%%%%
%%%%%%%%%%%%%%%%%%%%%
\def\ChapterStyle{\parindent=0pt\parskip=0pt\normalbaselineskip=15pt%
\def\nt{\fourteenSansSerifBold}%
\def\rm{\fam0\fourteenSansSerifBold}%
\textfont0=\fourteenSansSerifBold\scriptfont0=\tenSansSerifBold\scriptscriptfont0=\eightSansSerifBold%text(\fourteenrm)
\textfont1=\fourteenMath\scriptfont1=\tenMath\scriptscriptfont1=\eightMath%math(\fourteenmi)
\textfont2=\fourteenSymbols\scriptfont2=\tenSymbols\scriptscriptfont2=\eightSymbols%symbol(\fourteensy)
\textfont3=\fourteenMoreSymbols\scriptfont3=\tenMoreSymbols\scriptscriptfont3=\eightMoreSymbols%ex(fourteenex)
\textfont\itfam=\fourteenSansSerifItalics\def\it{\fam\itfam\fourteenSansSerifItalics}%
\textfont\slfam=\fourteenSansSerifSlanted\def\sl{\fam\slfam\fourteenSerifSansSlanted}%
\textfont\ttfam=\fourteenTypewriter\def\tt{\fam\ttfam\fourteenTypewriter}%
\textfont\bffam=\fourteenSansSerif%
\def\bf{\fam\bffam\fourteenSansSerif}\scriptfont\bffam=\tenSansSerif\scriptscriptfont\bffam=\eightSansSerif%
\def\cal{\fourteenSymbols}%
\def\greekbold{\fourteenMathBold}%
\def\gothic{\fourteenGothic}%
\def\Bbb{\fourteenDouble}%
\def\LieFont{\fourteenSerifItalics}%
\nt\normalbaselines\baselineskip=15pt%
}

%%%%%%%%%%%%%%%%%%%%%%%%%%%%%%%%%%%%%%%%%%%%%%%%%%%%%%%%
%%%%%%%%%%%%%%%%%%%%%
\def\SectionStyle{\parindent=0pt\parskip=0pt\normalbaselineskip=13pt%
\def\nt{\twelveSansSerifBold}%
\def\rm{\fam0\twelveSansSerifBold}%
\textfont0=\twelveSansSerifBold\scriptfont0=\eightSansSerifBold\scriptscriptfont0=\eightSansSerifBold%text(\fourteenrm)
\textfont1=\twelveMath\scriptfont1=\eightMath\scriptscriptfont1=\eightMath%math(\fourteenmi)
\textfont2=\twelveSymbols\scriptfont2=\eightSymbols\scriptscriptfont2=\eightSymbols%symbol(\fourteensy)
\textfont3=\twelveMoreSymbols\scriptfont3=\eightMoreSymbols\scriptscriptfont3=\eightMoreSymbols%ex(fourteenex)
\textfont\itfam=\twelveSansSerifItalics\def\it{\fam\itfam\twelveSansSerifItalics}%
\textfont\slfam=\twelveSansSerifSlanted\def\sl{\fam\slfam\twelveSerifSansSlanted}%
\textfont\ttfam=\twelveTypewriter\def\tt{\fam\ttfam\twelveTypewriter}%
\textfont\bffam=\twelveSansSerif%
\def\bf{\fam\bffam\twelveSansSerif}\scriptfont\bffam=\eightSansSerif\scriptscriptfont\bffam=\eightSansSerif%
\def\cal{\twelveSymbols}%
\def\bg{\twelveMathBold}%
\def\gothic{\twelveGothic}%
\def\Bbb{\twelveDouble}%
\def\LieFont{\twelveSerifItalics}%
\nt\normalbaselines\baselineskip=13pt%
}

%%%%%%%%%%%%%%%%%%%%%%%%%%%%%%%%%%%%%%%%%%%%%%%
\def\SubSectionStyle{\parindent=0pt\parskip=0pt\normalbaselineskip=13pt%
\def\nt{\twelveSansSerifItalics}%
\def\rm{\fam0\twelveSansSerifItalics}%
\textfont0=\twelveSansSerifItalics\scriptfont0=\eightSansSerifItalics\scriptscriptfont0=\eightSansSerifItalics%
\textfont1=\twelveMath\scriptfont1=\eightMath\scriptscriptfont1=\eightMath%
\textfont2=\twelveSymbols\scriptfont2=\eightSymbols\scriptscriptfont2=\eightSymbols%
\textfont3=\twelveMoreSymbols\scriptfont3=\eightMoreSymbols\scriptscriptfont3=\eightMoreSymbols%
\textfont\itfam=\twelveSansSerif\def\it{\fam\itfam\twelveSansSerif}%
\textfont\slfam=\twelveSansSerifSlanted\def\sl{\fam\slfam\twelveSerifSansSlanted}%
\textfont\ttfam=\twelveTypewriter\def\tt{\fam\ttfam\twelveTypewriter}%
\textfont\bffam=\twelveSansSerifBold%
\def\bf{\fam\bffam\twelveSansSerifBold}\scriptfont\bffam=\eightSansSerifBold\scriptscriptfont\bffam=\eightSansSerifBold%
\def\cal{\twelveSymbols}%
\def\greekbold{\twelveMathBold}%
\def\gothic{\twelveGothic}%
\def\Bbb{\twelveDouble}%
\def\LieFont{\twelveSerifItalics}%
\nt\normalbaselines\baselineskip=13pt%
}

%%%%%%%%%%%%%%%%%%%%%%%%%%%%%%%%%%%%%%%%%%%%%%%%%%%%%%%%%%%
%%%%%%%%%%%%%%%%%%
\def\AuthorStyle{\parindent=0pt\parskip=0pt\normalbaselineskip=14pt%
\def\nt{\tenSerif}%
\def\rm{\fam0\tenSerif}%
\textfont0=\tenSerif\scriptfont0=\sevenSerif\scriptscriptfont0=\fiveSerif%text(\tenrm)
\textfont1=\tenMath\scriptfont1=\sevenMath\scriptscriptfont1=\fiveMath%math(\tenmi)
\textfont2=\tenSymbols\scriptfont2=\sevenSymbols\scriptscriptfont2=\fiveSymbols%symbol(\tensy)
\textfont3=\tenMoreSymbols\scriptfont3=\sevenMoreSymbols\scriptscriptfont3=\fiveMoreSymbols%ex(tenex)
\textfont\itfam=\tenSerifItalics\def\it{\fam\itfam\tenSerifItalics}%
\textfont\slfam=\tenSerifSlanted\def\sl{\fam\slfam\tenSerifSlanted}%
\textfont\ttfam=\tenTypewriter\def\tt{\fam\ttfam\tenTypewriter}%
\textfont\bffam=\tenSerifBold%
\def\bf{\fam\bffam\tenSerifBold}\scriptfont\bffam=\sevenSerifBold\scriptscriptfont\bffam=\fiveSerifBold%
\def\cal{\tenSymbols}%
\def\greekbold{\tenMathBold}%
\def\gothic{\tenGothic}%
\def\Bbb{\tenDouble}%
\def\LieFont{\tenSerifItalics}%
\nt\normalbaselines\baselineskip=14pt%
}

%%%%%%%%%%%%%%%%%%%%%%%%%%%%%%%%%%%%%%%%%%%%%%%%%%%%%%%%%%%
%%%%%%%%%%%%%%%%%%
\def\AddressStyle{\parindent=0pt\parskip=0pt\normalbaselineskip=14pt%
\def\nt{\eightSerif}%
\def\rm{\fam0\eightSerif}%
\textfont0=\eightSerif\scriptfont0=\sevenSerif\scriptscriptfont0=\fiveSerif%text(\tenrm)
\textfont1=\eightMath\scriptfont1=\sevenMath\scriptscriptfont1=\fiveMath%math(\tenmi)
\textfont2=\eightSymbols\scriptfont2=\sevenSymbols\scriptscriptfont2=\fiveSymbols%symbol(\tensy)
\textfont3=\eightMoreSymbols\scriptfont3=\sevenMoreSymbols\scriptscriptfont3=\fiveMoreSymbols%ex(tenex)
\textfont\itfam=\eightSerifItalics\def\it{\fam\itfam\eightSerifItalics}%
\textfont\slfam=\eightSerifSlanted\def\sl{\fam\slfam\eightSerifSlanted}%
\textfont\ttfam=\eightTypewriter\def\tt{\fam\ttfam\eightTypewriter}%
\textfont\bffam=\eightSerifBold%
\def\bf{\fam\bffam\eightSerifBold}\scriptfont\bffam=\sevenSerifBold\scriptscriptfont\bffam=\fiveSerifBold%
\def\cal{\eightSymbols}%
\def\greekbold{\eightMathBold}%
\def\gothic{\eightGothic}%
\def\Bbb{\eightDouble}%
\def\LieFont{\eightSerifItalics}%
\nt\normalbaselines\baselineskip=14pt%
}

%%%%%%%%%%%%%%%%%%%%%%%%%%%%%%%%%%%%%%%%%%%%%%%%%%%%%%%%%%%
%%%%%%%%%%%%%%%%%%
\def\AbstractStyle{\parindent=0pt\parskip=0pt\normalbaselineskip=12pt%
\def\nt{\eightSerif}%
\def\rm{\fam0\eightSerif}%
\textfont0=\eightSerif\scriptfont0=\sevenSerif\scriptscriptfont0=\fiveSerif%text(\tenrm)
\textfont1=\eightMath\scriptfont1=\sevenMath\scriptscriptfont1=\fiveMath%math(\tenmi)
\textfont2=\eightSymbols\scriptfont2=\sevenSymbols\scriptscriptfont2=\fiveSymbols%symbol(\tensy)
\textfont3=\eightMoreSymbols\scriptfont3=\sevenMoreSymbols\scriptscriptfont3=\fiveMoreSymbols%ex(tenex)
\textfont\itfam=\eightSerifItalics\def\it{\fam\itfam\eightSerifItalics}%
\textfont\slfam=\eightSerifSlanted\def\sl{\fam\slfam\eightSerifSlanted}%
\textfont\ttfam=\eightTypewriter\def\tt{\fam\ttfam\eightTypewriter}%
\textfont\bffam=\eightSerifBold%
\def\bf{\fam\bffam\eightSerifBold}\scriptfont\bffam=\sevenSerifBold\scriptscriptfont\bffam=\fiveSerifBold%
\def\cal{\eightSymbols}%
\def\greekbold{\eightMathBold}%
\def\gothic{\eightGothic}%
\def\Bbb{\eightDouble}%
\def\LieFont{\eightSerifItalics}%
\nt\normalbaselines\baselineskip=12pt%
}

%%%%%%%%%%%%%%%%%%%%%%%%%%%%%%%%%%%%%%%%%%%%%
\def\RefsStyle{\parindent=0pt\parskip=0pt%
\def\nt{\eightSerif}%
\def\rm{\fam0\eightSerif}%
\textfont0=\eightSerif\scriptfont0=\sevenSerif\scriptscriptfont0=\fiveSerif%text(\tenrm)
\textfont1=\eightMath\scriptfont1=\sevenMath\scriptscriptfont1=\fiveMath%math(\tenmi)
\textfont2=\eightSymbols\scriptfont2=\sevenSymbols\scriptscriptfont2=\fiveSymbols%symbol(\tensy)
\textfont3=\eightMoreSymbols\scriptfont3=\sevenMoreSymbols\scriptscriptfont3=\fiveMoreSymbols%ex(tenex)
\textfont\itfam=\eightSerifItalics\def\it{\fam\itfam\eightSerifItalics}%
\textfont\slfam=\eightSerifSlanted\def\sl{\fam\slfam\eightSerifSlanted}%
\textfont\ttfam=\eightTypewriter\def\tt{\fam\ttfam\eightTypewriter}%
\textfont\bffam=\eightSerifBold%
\def\bf{\fam\bffam\eightSerifBold}\scriptfont\bffam=\sevenSerifBold\scriptscriptfont\bffam=\fiveSerifBold%
\def\cal{\eightSymbols}%
\def\greekbold{\eightMathBold}%
\def\gothic{\eightGothic}%
\def\Bbb{\eightDouble}%
\def\LieFont{\eightSerifItalics}%
\nt\normalbaselines\baselineskip=10pt%
}

%%%%%%%%%%%%%%%%%%%%%%%%%%%%%%%%%%%%%%%%%%%%%%%%%%%%%%%%%%%
%%%%%%%%%%%%%%%%%%%

%%%%%%%%%%%%%%%%%%%%%%%%%%%%%%%%%%%%%%%%%%%%%%%%%%%%%%%%%%%%
%%%%%%%%%%%%%%%%%%

%
%    Various Libraries.    Version 1.2.0.beta
%    The best use is to paste all of them into the papers
%     1/8/2005
%

%%%%%%%%%%%%%%%%%%%%%%%%%%%%%%
%%%%%%			Utilities		 %%%%%%
%%%%%%%%%%%%%%%%%%%%%%%%%%%%%%

% Definition modes  %
\def\ModeYes{yes}
\def\ModeNo{no}

\def\ModeUndef{undefined}

%%%%%%%%%%%%

\def\nx{\noexpand}
\def\ni{\noindent}
\def\newpage{\vfill\eject}

\def\ss{\vskip 5pt}
\def\ms{\vskip 10pt}
\def\bs{\vskip 20pt}

 \def\,{\mskip\thinmuskip}
 \def\!{\mskip-\thinmuskip}
 \def\>{\mskip\medmuskip}
 \def\;{\mskip\thickmuskip}

%%%%%%%%%%%%%%%%%%%%%%%%%%%%%%
%%%%%%		Bibliography		 %%%%%%
%%%%%%%%%%%%%%%%%%%%%%%%%%%%%%
%
% Usage:
%	[\SetModeAuto]
% ... 
%	\bib{libro1}{L.Fatibene, ...}
%	\bib{libro2}{L.Fatibene, ...}
% ...
%	(see \ref{libro2} and \ref{libro1})
% ...
% 	\ShowBiblio
%

% Definition modes  %
\def\refsModePost{post}
\def\refsModeAuto{auto}

\def\dbRefsSatusModeOk{ok}
\def\dbRefsSatusModeError{error}
\def\dbRefsSatusModeWarning{warning}

%%%%%%%%%%%%

\newcount\BNUM
\BNUM=0

\def\refs{}

\def\SetModePost{\xdef\refsMode{\refsModePost}}			%	Items are numbered by Citation order
		%	Items are numbered by Insertion order
\SetModePost

\def\dbRefsStatusOk{%
	\xdef\dbRefsStatus{\dbRefsSatusModeOk}%
	\xdef\dbRefsError{\ModeNo}%
	\xdef\dbRefsWarning{\ModeNo}%
	\xdef\dbRefsInfo{\ModeNo}%
}

\def\dbRefs{%
}

\def\dbRefsGet#1{%
	\xdef\found{N}\xdef\ikey{#1}\dbRefsStatusOk%
	\xdef\key{\ModeUndef}\xdef\tag{\ModeUndef}\xdef\tail{\ModeUndef}%
	\dbRefs%
}

\def\NextRefsTag{%
	\global\advance\BNUM by 1%
}
\def\ShowTag#1{{\bf [#1]}}

\def\dbRefsInsert#1#2{%
\dbRefsGet{#1}%
\if\found Y %
   \xdef\dbRefsStatus{\dbRefsSatusModeWarning}%
   \xdef\dbRefsWarning{record is already there}%
   \xdef\dbRefsInfo{record not inserted}%
\else%
   \toks2=\expandafter{\dbRefs}%
   \ifx\refsMode\refsModeAuto \NextRefsTag
    \xdef\dbRefs{%
   	\the\toks2 \nx\xdef\nx\dbx{#1}%
	\nx\ifx\nx\ikey %
		\nx\dbx\nx\xdef\nx\found{Y}%
		\nx\xdef\nx\key{#1}%
		\nx\xdef\nx\tag{\the\BNUM}%
		\nx\xdef\nx\tail{#2}%
	\nx\fi}%
	\global\xdef\refs{\refs \ss\ni[\the\BNUM]\ #2\par}%%%%
   \fi%   	
   \ifx\refsMode\refsModePost 
    \xdef\dbRefs{%
   	\the\toks2 \nx\xdef\nx\dbx{#1}%
	\nx\ifx\nx\ikey %
		\nx\dbx\nx\xdef\nx\found{Y}%
		\nx\xdef\nx\key{#1}%
		\nx\xdef\nx\tag{\ModeUndef}%
		\nx\xdef\nx\tail{#2}%
	\nx\fi}%
   \fi%
\fi%
}

\def\dbRefsEdit#1#2#3{\dbRefsGet{#1}%
\if\found N 
   \xdef\dbRefsStatus{\dbRefsSatusModeError}%
   \xdef\dbRefsError{record is not there}%
   \xdef\dbRefsInfo{record not edited}%
\else%
   \toks2=\expandafter{\dbRefs}%
   \xdef\dbRefs{\the\toks2%
   \nx\xdef\nx\dbx{#1}%
   \nx\ifx\nx\ikey\nx\dbx %
	\nx\xdef\nx\found{Y}%
	\nx\xdef\nx\key{#1}%
	\nx\xdef\nx\tag{#2}%
	\nx\xdef\nx\tail{#3}%
   \nx\fi}%
\fi%
}

\def\bib#1#2{\RefsStyle\dbRefsInsert{#1}{#2}%
	\ifx\dbRefsStatus\dbRefsSatusModeWarning %
		\message{^^J}%
		\message{WARNING: Reference [#1] is doubled.^^J}%
	\fi%
}

\def\ref#1{\dbRefsGet{#1}%
\ifx\found N %
  \message{^^J}%
  \message{ERROR: Reference [#1] unknown.^^J}%
  \ShowTag{??}%
\else%
	\ifx\tag\ModeUndef \NextRefsTag%
		\dbRefsEdit{#1}{\the\BNUM}{\tail}%
		\dbRefsGet{#1}%
		\global\xdef\refs{\refs \ss\ni [\tag]\ \tail\par}%%%%
	\fi
	\ShowTag{\tag}%
\fi%
}

\def\ShowBiblio{\ms\Ensure{\SectionEnsure}%
{\SectionStyle\ni References}%
{\RefsStyle\refs}%
}

%%%%%%%%%%%%%%%%%%%%%%%%%%%%%%
%%%%%%		Label DB			 %%%%%%
%%%%%%%%%%%%%%%%%%%%%%%%%%%%%%
\newcount\CHANGES
\CHANGES=0
\def\AuxFile{7}
\def\PreventDoubleOn{\xdef\PreventDoubleLabel{\ModeYes}}

\PreventDoubleOn

\def\StoreLabel#1#2{\xdef\itag{#2}% Mantiene FileAux e ritorna #2 in \itag
 \ifx\PreModeStatus\ModeNo %
   \message{^^J}%
   \errmessage{You can't use Check without starting with OpenPreMode (and finishing with ClosePreMode)^^J}%
 \else%
   \immediate\write\AuxFile{\nx\dbLabelPreInsert{#1}{\itag}}%     
   \dbLabelGet{#1}%
   \ifx\itag\tag %
   \else%
	\global\advance\CHANGES by 1%
 	\xdef\itag{(?.??)}%
    \fi%
   \fi%
}

\def\PreModeStatus{\ModeNo}

\def\ClosePreMode{\immediate\closeout\AuxFile%
  \ifnum\CHANGES=0%
	\message{^^J}%
	\message{**********************************^^J}%
	\message{**  NO CHANGES TO THE AuxFile  **^^J}%
	\message{**********************************^^J}%
 \else%
	\message{^^J}%
	\message{**************************************************^^J}%
	\message{**  PLAEASE TYPESET IT AGAIN (\the\CHANGES)  **^^J}%
    \errmessage{**************************************************^^ J}%
  \fi%
  \edef\PreModeStatus{\ModeNo}%
}

\def\dbLabelSatusModeOk{ok}

\def\dbLabelSatusModeWarning{warning}

\def\dbLabelStatusOk{%
	\xdef\dbLabelStatus{\dbLabelSatusModeOk}%
	\xdef\dbLabelError{\ModeNo}%
	\xdef\dbLabelWarning{\ModeNo}%
	\xdef\dbLabelInfo{\ModeNo}%
}

\def\dbLabel{%
}

\def\dbLabelGet#1{%
	\xdef\found{N}\xdef\ikey{#1}\dbLabelStatusOk%
	\xdef\key{\ModeUndef}\xdef\tag{\ModeUndef}\xdef\pre{\ModeUndef}%
	\dbLabel%
}

\def\ShowLabel#1{%
 \dbLabelGet{#1}%
 \ifx\tag \ModeUndef %
 	\global\advance\CHANGES by 1%
 	(?.??)%
 \else%
 	\tag%
 \fi%
}

\def\dbLabelPreInsert#1#2{\dbLabelGet{#1}%
\if\found Y %
  \xdef\dbLabelStatus{\dbLabelSatusModeWarning}%
   \xdef\dbLabelWarning{Label is already there}%
   \xdef\dbLabelInfo{Label not inserted}%
   \message{^^J}%
   \errmessage{Double pre definition of label [#1]^^J}%
\else%
   \toks2=\expandafter{\dbLabel}%
    \xdef\dbLabel{%
   	\the\toks2 \nx\xdef\nx\dbx{#1}%
	\nx\ifx\nx\ikey %
		\nx\dbx\nx\xdef\nx\found{Y}%
		\nx\xdef\nx\key{#1}%
		\nx\xdef\nx\tag{#2}%
		\nx\xdef\nx\pre{\ModeYes}%
	\nx\fi}%
\fi%
}

\def\dbLabelInsert#1#2{\dbLabelGet{#1}%
\xdef\itag{#2}%
\dbLabelGet{#1}%
\if\found Y %
	\ifx\tag\itag %
	\else%
	   \ifx\PreventDoubleLabel\ModeYes %
		\message{^^J}%
		\errmessage{Double definition of label [#1]^^J}%
	   \else%
		\message{^^J}%
		\message{Double definition of label [#1]^^J}%
	   \fi%	
	\fi%
   \xdef\dbLabelStatus{\dbLabelSatusModeWarning}%
   \xdef\dbLabelWarning{Label is already there}%
   \xdef\dbLabelInfo{Label not inserted}%
\else%
   \toks2=\expandafter{\dbLabel}%
    \xdef\dbLabel{%
   	\the\toks2 \nx\xdef\nx\dbx{#1}%
	\nx\ifx\nx\ikey %
		\nx\dbx\nx\xdef\nx\found{Y}%
		\nx\xdef\nx\key{#1}%
		\nx\xdef\nx\tag{#2}%
		\nx\xdef\nx\pre{\ModeNo}%
	\nx\fi}%
\fi%
}

%%%%%%%%%%%%%%%%%%%%%%%%%%%%%%
%%%%%%		Numbering			 %%%%%%
%%%%%%%%%%%%%%%%%%%%%%%%%%%%%%

\newcount\PART
\newcount\CHAPTER
\newcount\SECTION
\newcount\SUBSECTION
\newcount\FNUMBER
%\newdimen\TOBOTTOM
%\newdimen\LIMIT

\PART=0
\CHAPTER=0
\SECTION=0
\SUBSECTION=0	
\FNUMBER=0

\def\LastPart{\ModeUndef}
\def\LastChapter{\ModeUndef}
\def\LastSection{\ModeUndef}
\def\LastSubSection{\ModeUndef}
\def\LastClaim{\ModeUndef}
\def\Last{\ModeUndef}

\newdimen\TOBOTTOM
\newdimen\LIMIT

\def\Ensure#1{\ \par\ \immediate\LIMIT=#1\immediate\TOBOTTOM=\the\pagegoal\advance\TOBOTTOM by -\pagetotal%
\ifdim\TOBOTTOM<\LIMIT\newpage \else%
\vskip-\parskip\vskip-\parskip\vskip-\baselineskip\fi}

%%%%%%%%%%%%%%%%%%%%%%%%%%%%%
\def\PartLabel{\the\PART}
\def\NewPart#1{\global\advance\PART by 1%
         \bs\ni{\PartStyle  Part \PartLabel:}
         \bs\ni{\PartStyle #1}\newpage%
         \CHAPTER=0\SECTION=0\SUBSECTION=0\FNUMBER=0%
         \gdef\Left{#1}%
         \global\edef\Last{\PartLabel}%
         \global\edef\LastPart{\PartLabel}%
         \global\edef\LastChapter{\ModeUndef}%
         \global\edef\LastSection{\ModeUndef}%
         \global\edef\LastSubSection{\ModeUndef}%
         \global\edef\LastClaim{\ModeUndef}}
%%%%%%%%%%%%%%%%%%%%%%%%%%%%%
\def\ChapterLabel{\the\CHAPTER}
\def\NewChapter#1{\global\advance\CHAPTER by 1%
         \bs\ni{\ChapterStyle  Chapter \ChapterLabel: #1}\ms%
         \SECTION=0\SUBSECTION=0\FNUMBER=0%
         \gdef\Left{#1}%
         \global\edef\Last{\ChapterLabel}%
         \global\edef\LastChapter{\ChapterLabel}%
         \global\edef\LastSection{\ModeUndef}%
         \global\edef\LastSubSection{\ModeUndef}%
         \global\edef\LastClaim{\ModeUndef}}
%%%%%%%%%%%%%%%%%%%%%%%%%%%%%
\def\SectionEnsure{3cm}
\def\NewSection#1{\Ensure{\SectionEnsure}\gdef\SectionLabel{\the\SECTION}\global\advance\SECTION by 1%
         \ms\ni{\SectionStyle  \SectionLabel.\ #1}\ss%
         \SUBSECTION=0\FNUMBER=0%
         \gdef\Left{#1}%
         \global\edef\Last{\SectionLabel}%
         \global\edef\LastSection{\SectionLabel}%
         \global\edef\LastSubSection{\ModeUndef}%
         \global\edef\LastClaim{\ModeUndef}}
%%%%%%%%%%%%%%%%%%%%%%%%%%%%%
\def\NewAppendix#1#2{\Ensure{\SectionEnsure}\gdef\SectionLabel{#1}\global\advance\SECTION by 1%
         \bs\ni{\SectionStyle  Appendix \SectionLabel.\ #2}\ss%
         \SUBSECTION=0\FNUMBER=0%
         \gdef\Left{#2}%
         \global\edef\Last{\SectionLabel}%
         \global\edef\LastSection{\SectionLabel}%
         \global\edef\LastSubSection{\ModeUndef}%
         \global\edef\LastClaim{\ModeUndef}}
%%%%%%%%%%%%%%%%%%%%%%%%%%%%%
\def\Acknowledgements{\Ensure{\SectionEnsure}\gdef\SectionLabel{}%
         \ms\ni{\SectionStyle  Acknowledgments}\ss%
         \SECTION=0\SUBSECTION=0\FNUMBER=0%
         \gdef\Left{}%
         \global\edef\Last{\ModeUndef}%
         \global\edef\LastSection{\ModeUndef}%
         \global\edef\LastSubSection{\ModeUndef}%
         \global\edef\LastClaim{\ModeUndef}}
%%%%%%%%%%%%%%%%%%%%%%%%%%%%%
\def\SubSectionEnsure{2cm}
\def\SubSectionLabel{\ifnum\SECTION>0 \the\SECTION.\fi\the\SUBSECTION}
\def\NewSubSection#1{\Ensure{\SubSectionEnsure}\global\advance\SUBSECTION by 1%
         \ms\ni{\SubSectionStyle #1}\ss%
         \global\edef\Last{\SubSectionLabel}%
         \global\edef\LastSubSection{\SubSectionLabel}}
%%%%%%%%%%%%%%%%%%%%%%%%%%%%%
\def\SetNumberingModeN{\def\ClaimLabel{(\the\FNUMBER)}}
\def\SetNumberingModeSN{\def\ClaimLabel{(\ifnum\SECTION>0 \SectionLabel.\fi%
      \the\FNUMBER)}}
\def\SetNumberingModeCSN{\def\ClaimLabel{(\ifnum\CHAPTER>0 \the\CHAPTER.\fi%
      \ifnum\SECTION>0 \SectionLabel.\fi%
      \the\FNUMBER)}}

\def\NewClaim{\global\advance\FNUMBER by 1%
    \ClaimLabel%
    \global\edef\LastClaim{\ClaimLabel}%
    \global\edef\Last{\ClaimLabel}}
%%%%%%%%%%%%%%%%%%%%%%%%%%%%%

\def\HideLabels{\xdef\ShowLabelsMode{\ModeNo}}
\HideLabels

\def\fn{\eqno{\NewClaim}} 
\def\fl#1{%
\ifx\ShowLabelsMode\ModeYes%
%\eqno{\relax\hbox to 1cm{\NewClaim\hbox{[#1]}}}%
 \eqno{{\buildrel{\hbox{\AbstractStyle[#1]}}\over{\hfill\NewClaim}}}%
\else%
 \eqno{\NewClaim}%
\fi% 
\dbLabelInsert{#1}{\ClaimLabel}}
\def\fprel#1{\global\advance\FNUMBER by 1\StoreLabel{#1}{\ClaimLabel}%
\ifx\ShowLabelsMode\ModeYes%
%\eqno{\relax\hbox to 1cm{ .\itag\hbox{[#1]}}}%
\eqno{{\buildrel{\hbox{\AbstractStyle[#1]}}\over{\hfill.\itag}}}%
\else%
 \eqno{\itag}%
\fi% 
}

\def\cl#1{\global\advance\FNUMBER by 1\dbLabelInsert{#1}{\ClaimLabel}%
\ifx\ShowLabelsMode\ModeYes%
${\buildrel{\hbox{\AbstractStyle[#1]}}\over{\hfill\ClaimLabel}}$%
\else%
  $\ClaimLabel$%
\fi% 
}
\def\cprel#1{\global\advance\FNUMBER by 1\StoreLabel{#1}{\ClaimLabel}%
\ifx\ShowLabelsMode\ModeYes%
${\buildrel{\hbox{\AbstractStyle[#1]}}\over{\hfill.\itag}}$%
\else%
  $\itag$%
\fi% 
}
%%%%%%%%%%%%%%%%%%%%%%%%%%%%%

\def\Note{\ms\leftskip 3cm\rightskip 1.5cm\AbstractStyle}
\def\endNote{\par\leftskip 2cm\rightskip 0cm\NormalStyle\ss}

%%%%%%%%%%%%%%%%%%%%%%%%%%%%%%
%%%%%%		   Sidebars		          %%%%%%
%%%%%%%%%%%%%%%%%%%%%%%%%%%%%%

\parindent=7pt
\leftskip=2cm
\newcount\SideIndent
\newcount\SideIndentTemp
\SideIndent=0
\newdimen\SectionIndent
\SectionIndent=-8pt

\def\sidebar{\vrule height15pt width.2pt }
\def\endcorner{\hbox{\hbox{\vrule height6pt width.2pt}\vbox to6pt{\vfill\hbox
to4pt{\leaders\hrule height0.2pt\hfill}}}}
\def\begincorner{\hbox{\hbox{\vrule height6pt width.2pt}\vbox to6pt{\hbox
to4pt{\leaders\hrule height0.2pt\hfill}}}}
\def\endbegincorner{\hbox{\vbox to15pt{\endcorner\vskip-6pt\begincorner\vfill}}}
\def\SideShow{\SideIndentTemp=\SideIndent \ifnum \SideIndentTemp>0 
\loop\sidebar\hskip 2pt \advance\SideIndentTemp by-1\ifnum \SideIndentTemp>1 \repeat\fi}

\def\BeginSection{{\vbadness 100000 \par\ni\hskip\SectionIndent%
\SideShow\vbox to 15pt{\vfill\begincorner}}\global\advance\SideIndent by1\vskip-10pt}

\def\EndSection{{\vbadness 100000 \par\ni\global\advance\SideIndent by-1%
\hskip\SectionIndent\SideShow\vbox to15pt{\endcorner\vfill}\vskip-10pt}}

\def\EndBeginSection{{\vbadness 100000\par\ni%
\global\advance\SideIndent by-1\hskip\SectionIndent\SideShow
\vbox to15pt{\vfill\endbegincorner}}%
\global\advance\SideIndent by1\vskip-10pt}

\def\ShowBeginCorners#1{%
\SideIndentTemp =#1 \advance\SideIndentTemp by-1%
\ifnum \SideIndentTemp>0 %
\vskip-15truept\hbox{\kern 2truept\vbox{\hbox{\begincorner}%
\ShowBeginCorners{\SideIndentTemp}\vskip-3truept}}%				
\fi%
}

\def\ShowEndCorners#1{%
\SideIndentTemp =#1 \advance\SideIndentTemp by-1%
\ifnum \SideIndentTemp>0 %
\vskip-15truept\hbox{\kern 2truept\vbox{\hbox{\endcorner}%
\ShowEndCorners{\SideIndentTemp}\vskip 2truept}}%				
\fi%
}

\def\BeginSections#1{{\vbadness 100000 \par\ni\hskip\SectionIndent%
\SideShow\vbox to 15pt{\vfill\ShowBeginCorners{#1}}}\global\advance\SideIndent by#1\vskip-10pt}

\def\EndSections#1{{\vbadness 100000 \par\ni\global\advance\SideIndent by-#1%
\hskip\SectionIndent\SideShow\vbox to15pt{\vskip15pt\ShowEndCorners{#1}\vfill}\vskip-10pt}}

\def\EndBeginSections#1#2{{\vbadness 100000\par\ni%
\global\advance\SideIndent by-#1%
\hbox{\hskip\SectionIndent\SideShow\kern-2pt%
\vbox to15pt{\vskip15pt\ShowEndCorners{#1}\vskip4pt\ShowBeginCorners{#2}}}}%
\global\advance\SideIndent by#2\vskip-10pt}

%%%%%%%%%%%%%%%%%%%%%%%%%%%%%%
%%%%%%		Margin notes		 %%%%%%
%%%%%%%%%%%%%%%%%%%%%%%%%%%%%%

%%%%%%%%%%%%%%%%%%%%%%%%%%%%%

%%%%%%%%%%%%%%%%%%%%%%%%%%%%%

%
%    Macros.    Version 1.2.0.beta
%    The best use is to paste all of them into the papers
%     1/8/2005
%

%%%%%%%%%%%%%%%%%%%%%%%%%%%%%%
%%%%%%			Greek		 %%%%%%
%%%%%%%%%%%%%%%%%%%%%%%%%%%%%%

\def\al{\alpha}
\def\be{\beta}
\def\de{\delta}
\def\ga{\gamma}

\def\ep{\epsilon}

\def\la{\lambda}
\def\ze{\zeta}
\def\om{\omega}
\def\si{\sigma}

\def\Ga{\Gamma}

\def\La{\Lambda}
\def\Om{\Omega}
\def\Si{\Sigma}

%%%%%%%%%%%%%%%%%%%%%%%%%%%%%%
%%%%%%			Cal			 %%%%%%
%%%%%%%%%%%%%%%%%%%%%%%%%%%%%%

 \def\calE{{\hbox{\cal E}}}

%%%%%%%%%%%%%%%%%%%%%%%%%%%%%%
%%%%%%			gothic		 %%%%%%
%%%%%%%%%%%%%%%%%%%%%%%%%%%%%%

 		% to prevent \sl redefinition

%%%%%%%%%%%%%%%%%%%%%%%%%%%%%%
%%%%%%			Bbb			 %%%%%%
%%%%%%%%%%%%%%%%%%%%%%%%%%%%%%
 \def\one{{\hbox{\Bbb I}}}
 
 \def\R{{\hbox{\Bbb R}}}

 \def\R{{\hbox{\Bbb R}}}

%%%%%%%%%%%%%%%%%%%%%%%%%%%%%%
%%%%%%		MathRoman		 %%%%%%
%%%%%%%%%%%%%%%%%%%%%%%%%%%%%%

\def\SO{{\hbox{SO}}}

\def\GL{{\hbox{GL}}}
\def\det{{\hbox{det}}}

\def\id{{\hbox{\rm id}}}

%%%%%%%%%%%%%%%%%%%%%%%%%%%%%%
%%%%%%		OtherSymbols		 %%%%%%
%%%%%%%%%%%%%%%%%%%%%%%%%%%%%%
\def\ip{\hbox to4pt{\leaders\hrule height0.3pt\hfill}\vbox to8pt{\leaders\vrule width0.3pt\vfill}\kern 2pt}
% inner product
 
\def\del{\partial}
\def\na{\nabla}

\def\Lie{\hbox{\LieFont \$}}

\def\arr{\rightarrow}

%
%    Format.    Version 1.2.0.beta
%    The best use is to paste all of them into the papers
%     1/8/2005
%

\def\cases#1{\left\{\eqalign{#1}\right.}
%%%%%%%%%%%%%%%%%%%%
\NormalStyle
\SetNumberingModeSN
\PreventDoubleOn

\long\def\title#1{\centerline{\TitleStyle\ni#1}}

\long\def\author#1{\ms\centerline{\AuthorStyle by {\it #1}}}

\long\def\address#1{\ss\centerline{\AddressStyle #1}\par}
\long\def\moreaddress#1{\centerline{\AddressStyle #1}\par}
\def\abstract{\ms\leftskip 3cm\rightskip .5cm\AbstractStyle{\bf \ni Abstract:}\ }
\def\endabstract{\par\leftskip 2cm\rightskip 0cm\NormalStyle\ss}

%%%%%%%%%%%%%%%%%%%%%%%%%%%%%
\SetNumberingModeSN
%\ShowLabels

\def\frac[#1/#2]{\hbox{$#1\over#2$}}
\def\Frac[#1/#2]{{#1\over#2}}
\def\({\left(}
\def\){\right)}
\def\[{\left[}
\def\]{\right]}
\def\^#1{{}^{#1}_{\>\cdot}}
\def\_#1{{}_{#1}^{\>\cdot}}
\def\Label=#1{{\buildrel {\hbox{\fiveSerif \ShowLabel{#1}}}\over =}}
\def\<{\kern -1pt}

%%%%%%%%  		Collapsable Notes		%%%%%%%%%%%%%%%%%%%%%%%%

\def\ExpandAllCNotes{\long\def\CNote##1{%
\BeginSection%\Margine{{\AbstractStyle To be collapsed}}%
	\Note%
 		##1%
	\endNote% 
\EndSection%
}}
\ExpandAllCNotes
%
% If you want to collapse classes of CNotes independently one of the other, just clone the definition as
%
%	\def\CollapseAllCNotesClassA{\long\def\CNoteClassA##1{}}
%	\def\ExpandAllCNotesClassA{\long\def\CNoteClassA##1{\BeginSection\Note ##1 \endNote\EndSection}}
%	\ExpandAllCNotesClassA
%
%	\def\CollapseAllCNotesClassB{\long\def\CNoteClassB##1{}}
%	\def\ExpandAllCNotesClassB{\long\def\CNoteClassB##1{\BeginSection\Note ##1 \endNote\EndSection}}
%	\ExpandAllCNotesClassB
%
%%%%%%%%%%%%%%%%%%%%%%%%%%%%%%%%%%%%%%%%%%%%%%%%%

%%%%%%%%%%%%			frames 				%%%%%%%%%%%%%%%%%%%

\def\frame#1{\vbox{\hrule\hbox{\vrule\vbox{\kern2pt\hbox{\kern2pt#1\kern2pt}\kern2pt}\vrule}\hrule\kern-4pt}}

\def\Box to #1#2#3{\frame{\vtop{\hbox to #1{\hfill #2 \hfill}\hbox to #1{\hfill #3 \hfill}}}}

%%%%%%%%%%%%%%%%%%%%%%%%%%%%%%%%%%%%%%%%%%%%%%%%%

%%%%%%%%%%%%%%%%%%%%%%%%%%%%%%%%%%%%%%%%%%%%%%%%%%%

\def\ubal{\underline{\al}\kern1pt}
\def\obal{\overline{\al}\kern1pt}

\def\ubR{\underline{R}\kern1pt}
\def\obR{\overline{R}\kern1pt}
\def\ubom{\underline{\om}\kern1pt}
\def\obxi{\overline{\xi}\kern1pt}
\def\ubu{\underline{u}\kern1pt}
\def\ube{\underline{e}\kern1pt}
\def\obe{\overline{e}\kern1pt}

\bib{Ortin}{T.\ Ortin,
%{\it A Note on Lie-Lorentz Derivatives},
Class. Quant. Grav. {\bf 19} (2002) L143; %-L150; 
hep-th/0206159
}

\bib{Spinors}{L.\ Fatibene, M.\ Ferraris,  M.\ Francaviglia, M.\ Godina,
Gen.\ Rel.\ Grav.\ {\bf 30} (9) (1998) 1371.%--1389.
}

\bib{Godina}{L. Fatibene, M. Ferraris, M. Francaviglia, M. Godina,
%{\it A geometric definition of Lie derivative for Spinor Fields},
in: Proceedings of {\it ``6th International Conference on Differential Geometry
and its Applications, August 28--September 1, 1995"}, (Brno, Czech Republic),
Editor: I. Kol{\'a}{\v r}, MU University, Brno, Czech Republic (1996) 549.}

\bib{Kosman}{Y. Kosmann, Ann. di Matematica Pura e Appl. {\bf 91} (1971) 317.%--395.
\goodbreak
Y. Kosmann, Comptes Rendus Acad. Sc. Paris, s\'erie A, {\bf 262} (1966) 289.%--292.
\goodbreak
Y. Kosmann, Comptes Rendus Acad. Sc. Paris, s\'erie A, {\bf 262} (1966) 394.%--397.
\goodbreak
Y. Kosmann, Comptes Rendus Acad. Sc. Paris, s\'erie A, {\bf 264} (1967) 355.%--358.
}

\bib{Ray1}{L.\ Fatibene, M.\ Ferraris, M.\ Francaviglia, R.G.\ McLenaghan,
%{\it Generalized symmetries in mechanics and field theories'}, 
J. Math. Phys. {\bf 43} (2002), 3147.%-3161
}

\bib{Ray2}{L.\ Fatibene, R.G.\ McLenaghan, S.\ Smith,
%{\it Separation of variables for the Dirac equation on low dimensional spaces},
in: Advances in general relativity and cosmology, Pitagora, Bologna (2003) 109.%-127
}

\bib{Trautman}{A.\ Trautman, 
%{\it Invariance of Lagrangian Systems},
in: ``Papers in honour of J. L. Synge'',
Clarenden Press, Oxford, (1972) 85.
}

\bib{Kolar}{I. Kol{\'a}{\v r}, P.W. Michor, J. Slov{\'a}k,
{\it Natural Operations in Differential Geometry},
(Springer--Verlag, N.Y., 1993)}

\bib{Libro}{L. Fatibene, M. Francaviglia,
{\it Natural and Gauge Natural Formalism for Classical Field Theories},
Kluwer Academic Publishers, (Dordrecht, 2003), xxii
}

\bib{Jadwisin}{L. Fatibene, M. Francaviglia, 
%{\it Deformations of spin structures and gravity},
Acta Physica Polonica B, {\bf 29} (4) (1998) 915.%-928.
}

\bib{Yano}{K. Yano, 
{\it The theory of Lie derivatives and its applications}, 
North-Holland, (Amsterdam, 1955)}

\bib{Sharipov}{R. Sharipov,
{\it A note on Kosmann-Lie derivatives of Weyl spinors}, arxiv: 0801.0622}

\bib{Obukhov}{Y.N. Obukhov, G.F. Rubilar, Phys. Rev. D {\bf 74}, (2006) 064002; gr-qc/0608064},

\bib{Vandyck2}{M. A. Vandyck, Gen. Rel. Grav. {\bf 20} (1988) 261. }
\bib{Vandyck3}{M. A. Vandyck, Gen. Rel. Grav. {\bf 20} (1988) 905.} 
\bib{Mtheo}{J. M. Figueroa-OÕFarrill, Class. Quant. Grav. {\bf 16} (1999) 2043; hep-th/9902066. }
\bib{Vandyck1}{D. J. Hurley and M. A. Vandyck, J. Phys. A {\bf 27} (1994) 4569. }

\bib{Bou}{J.-P. Bourguignon, P. Gauduchon, 
%{\it Spineurs, opŽrateurs de Dirac et variations de mŽtriques }, 
Commun. Math. Phys. {\bf 144} (1992), 581.%--599
}

\bib{Holst}{S.\ Holst, 
%{\it Barbero's Hamiltonian Derived from a Generalized Hilbert-Palatini Action},
Phys.\ Rev.\ {\bf D53} (1996) 5966.}

\bib{Barbero}{F.\ Barbero, 
%{\it Real Ashtekar variables for Lorentzian signature space-time},
Phys.\ Rev.\ {\it D51} (1996), 5507.}

\bib{Immirzi}{G.\ Immirzi, 
%{\it Quantum Gravity and Regge Calculus},
Nucl.\ Phys.\ Proc.\ Suppl.\ {\bf 57}, (1997) 65.%--72
}

\bib{myBI}{L.\ Fatibene, M.\ Francaviglia, C.\ Rovelli, 
%{\it On a Covariant Formulation of the Barbero-Immirzi Connection}, 
Class. Quantum Grav. 24 (2007) 3055.%--3066.
}

\bib{Rov2}{L. Fatibene, M.Francaviglia, C.Rovelli, 
%{\it Lagrangian Formulation of Ashtekar-Barbero-Immirzi Gravity}
CQG {\bf 24} (2007) 4207.%--4217
}

\bib{GoMa}{M. Godina, P. Matteucci,
%{\it The Lie derivative of spinor fields: theory and applications},
Int. J. Geom. Methods Mod. Phys. {\bf 2} (2005) 159; %--188; 
math/0504366}

%%%%%%%%%%%%%%%%%%%%%%%%%%%%%%%%%%%%%%%%%%%%%%%%%%%
\NormalStyle
%\ShowLabels
%\CollapseAllCNotes

\title{General theory of Lie derivatives for Lorentz tensors\footnote{$^\ast$}{{\AbstractStyle
	This paper is published despite the effects of the Italian law 133/08 ({\tt http://groups.google.it/group/scienceaction}). 
        This law drastically reduces public funds to public Italian universities, which is particularly dangerous for free scientific research, 
        and it will prevent young researchers from getting a position, either temporary or tenured, in Italy.
        The authors are protesting against this law to obtain its cancellation.\goodbreak}}}

\author{L.Fatibene$^{a, b}$, M.Francaviglia$^{a,b, c}$\footnote{}{\AbstractStyle eMail: {\tt lorenzo.fatibene@unito.it}, {\tt mauro.francaviglia@unito.it}}}

\address{$^a$ Department of Mathematics, University of Torino (Italy)}

\moreaddress{$^b$ INFN - Torino: Iniziativa Specifica NA12}

\moreaddress{$^c$ LCS, University of Calabria}



%\maketitle

\abstract
We show how the {\it ad hoc} prescriptions appeared in 2001 for the Lie derivative of Lorentz tensors (see
\ref{Ortin}) are a direct consequence of the Kosmann lift defined earlier (1996, see \ref{Godina}), in a much
more general setting encompassing older results of Y.~Kosmann (1971, see \ref{Kosman}).

\endabstract

\NewSection{Introduction}

The geometric theory of Lie derivatives of spinor fields is an old and intriguing issue that is
relevant in many contexts, among which we quote the applications in Supersymmetry (see \ref{Ortin}, \ref{Ray1}) and the problem of
separation of variables of Dirac equation (see \ref{Ray2}). It is as well essential for the understanding of the
general foundations of the theory of spinor fields and, eventually, of General Relativity as a whole. We stress
that despite spinor fields can be endowed with a correct physical interpretation only in a quantum  framework,
this quantum field theory is obtained by quantization procedures from a classical variational problem. Hence even
if a classical field theory describing spinors is not endowed with a direct physical interpretation its
variational issues (field equations and conserved quantities) are mathematically interesting on
their own as well as they have important consequences on the corresponding quantum field theory.

The situation in Minkowski spacetime (as well as on other maximally symmetric spaces) is pretty well established
and it is based on the existence of sufficiently many Killing vectors $\xi$.
The problem of Lie derivatives arises when one wants to generalize these arguments to 
more general spacetimes, i.e.~when Killing vectors are less than enough, or when coupling with gravity, i.e.~when 
the metric background cannot be regarded as being fixed {\it a priori} but it has to be determined
dynamically by field equations.
A definition for Lie derivatives of spinors along generic spacetime vector fields,  not necessarily Killing ones,
on a general curved spacetime was already proposed in 1971 by Y.~Kosmann \ref{Kosman} by an {\it ad
hoc} prescription. 
In 1996 we and coauthors (see also \ref{GoMa}) provided a geometric framework which justifies the {\it ad hoc} prescription within the general
framework of Lie derivatives on fiber bundles (see also \ref{Trautman}, \ref{Sharipov} and \ref{Bou}) in the explicit context of gauge natural
bundles \ref{Kolar} which turn out to be the most appropriate arena for (gauge-covariant) field theories \ref{Libro}.

The key point is the construction of the (generalized) {\it Kosmann lift} (so-called by us in honour of the original
{\it ad hoc} prescription) which is induced by any spacetime frame. 
This lift is defined on any principal bundle $\Si$ having  the special orthogonal group as structure group in any dimension and
signature. According to this prescription a spacetime vector field $\xi$ is uniquely lifted to
a bundle vector field $\hat\xi_\Si$.

This lift $\hat\xi_\Si$ on the principal bundle $\Si$ defines in turn the Lie derivative operator on sections of {\it any}
fiber bundle associated  to $\Si$, where objects like spinors or spin--connections  are defined as sections. 
Unfortunately, this Lie derivative is not {\it natural}, in the sense that it does not
preserve the commutator unless it is restricted to Killing vectors only. 
However, we stress that an advantage of this framework consists in showing and definitely explaining why there cannot be and in fact there is no possible
natural prescription for the Lie derivative of spinors. As a consequence, one has to choose whether to restrict
artificially to Killing vectors (which is certainly physically impossible unless under extremely special
conditions) or to learn how to cope with the fact that spinors are non-natural objects. The gauge natural
formalism is a possible escape (see \ref{Spinors}). In any case unless restricting to very special situation,
one has to define Lie derivatives with respect to arbitrary spacetime vector fields. Furthermore, even in special
situations one can {\it a posteriori} restrict the vector field to be Killing one (if any exists) in order to obtain a unifying view
on the matter, in which all Lie derivatives are obtained as a specialization of a general notion.  

The very same framework introduced for spinors provides a suitable arena to deal with {\it Lorentz tensors} in GR.
Similar approaches can be found in the literature (see \ref{Yano}) as well as more recently (see \ref{Obukhov}).
In GR there are many objects which are endowed with specific transformation rules with respect to Lorentz transformations,
even though, of course, in GR these transformations cannot be implemented in general by a subgroup of the whole group of all diffeomorphisms.
Let us mention e.g.~tetrads and spin connections in a Cartan framework, where pointwise Lorentz transformations act as a gauge group.
This framework is also the kinematical arena to define the self-dual formulation of GR that is the starting point of LQG approach.

We shall here review the general theory of Lorentz tensors and their Lie derivative and compare with
the direct and {\it ad hoc} method based on Killing vectors appeared in \ref{Ortin}.
 The key issue consists in recognizing that Lorentz tensors are, {\it by definition}, sections of some
bundle associated to a suitable principal bundle $\Si$  by means of the appropriate tensorial representation of the appropriate special orthogonal structure group.

\NewSection{The Kosmann lift}

Let $M$ be a $m$--dimensional manifold (which will be required to allow global metrics of signature $\eta=(r,s)$, with $m=r+s$).
Let us denote by $x^\mu$ local coordinates on $M$, which induce a basis $\del_\mu$ of tangent spaces;
let $L(M)$ denote the {\it general frame bundle} of $M$ and set $(x^\mu, V_a^\mu)$ for fibered coordinates on $L(M)$.
We can define a right--invariant basis for vertical vectors on $L(M)$
$$
\rho^\mu_\nu= V^\mu_a \Frac[\del/\del V^\nu_a]
\fn$$
The general frame bundle is natural (see \ref{Kolar}), hence any spacetime vector field $\xi=\xi^\mu\del_\mu$ defines a natural lift on $L(M)$
$$
\hat\xi=\xi^\mu\>\del_\mu+ \del_\mu\xi^\nu \>\rho^\mu_\nu
\fn$$
We stress that the lift vector field $\hat \xi$ is global whenever $\xi$ is global.

A connection on $L(M)$ is denoted  by $\Ga^\al_{\be\mu}$ and it defines a lift
$$
\Ga:TM\arr TL(M):\xi^\mu\del_\mu\mapsto  \xi^\mu \( \del_\mu - \Ga^\al_{\be\mu} \rho^\be_\al\)
\fn$$
This lift does not in general preserve commutators, unless the connection is flat.

Ordinary tensors are sections of bundles associated to $L(M)$. 
The connection $\Ga^\al_{\be\mu}$ induces connections on associated bundles and defines in turn the covariant derivatives of ordinary tensors.

\Note
For example, tensors of rank $(1,1)$ are sections of the bundle $T^1_1(M)$ associated to $L(M)$ using the appropriate tensor representations, namely
$$
\la:\GL(m)\times V\arr V: (J^\mu_\nu, t^\mu_\nu)\mapsto  t'^\mu_\nu= J^\mu_\al t^\al_\be  \bar J^\be_\nu
\fn$$
where the bar denotes the inverse in $\GL(n, \R)$.

The connection $\Ga$ on $L(M)$ induces on this associated bundle the connection
$$
T^1_1(\Ga)= dx^\mu \otimes \(\del_\mu - \(\Ga^\al_{\ga\mu} t^\ga_\be - \Ga^\ga_{\be\mu} t^\al_\ga\)\frac[\del/\del t^\al_\be]\)
\fn$$
which in turn defines the standard covariant derivative of such tensors:
$$
\na_\xi t= Tt (\xi) - T^1_1(\Ga)(\xi)= \(d_\mu t^\al_\be + \Ga^\al_{\ga\mu} t^\ga_\be - \Ga^\ga_{\be\mu} t^\al_\ga\)\Frac[\del/\del t^\al_\be]
\fn$$
\endNote

If a metric $g=g_{\mu\nu} \>dx^\mu\otimes dx^\nu$ is given on $M$ then its Christoffel symbols define the Levi-Civita connection of the metric.
Such a connection is torsionless (i.e.~symmetric in lower indices) and {\it compatible with the metric}, i.e.~such that $\na_\mu g_{\al\be}=0$.

Let now $(\Si, M, \pi, \SO(\eta))$ be a principal bundle over the manifold $M$ and
let $(x^\mu, S^a_b)$ be (overdetermined) fibered  ``coordinates'' on the principal bundle $\Si$.
We can define a right--invariant  pointwise basis $\si_{ab}$ for vertical vectors on $\Si$ by setting
$$
\si_{ab}= \eta_{d[a} \rho^d_{b]}
\qquad\qquad
\rho^d_{b}= S^d_c {\del\over\del S^{b}_c}
\fn$$
where $\eta_{ab}$ is the canonical diagonal matrix of signature $\eta=(r,s)$ and square brackets denote skew-symmetrization over indices.

A connection on $\Si$ is in the form
$$
\om= dx^\mu \otimes \(\del_\mu -\om^{ab}_{\mu} \si_{ab}\)
\fn$$
Also in this case the connection on $\Si$ induces connections on any associated bundle and there defines covariant derivatives of sections.

A {\it frame} is a bundle  map $e:\Si\arr L(M)$ which preserves the right action, i.e.~such that
$$
\begindc{\commdiag}[1]
\obj(0,80)[Si]{$\Si$}
\obj(80,80)[LM]{$L(M)$}
\obj(0,20)[M1]{$M$}
\obj(80,20)[M2]{$M$}
\mor{Si}{M1}{}
\mor{LM}{M2}{}
\mor{Si}{LM}{$e$}
\mor{M1}{M2}{}[\atleft, \solidline] \mor(0,23)(80,23){}[\atleft, \solidline]
\enddc
\qquad\qquad\qquad
\begindc{\commdiag}[1]
\obj(0,80)[Si]{$\Si$}
\obj(80,80)[LM]{$L(M)$}
\obj(0,20)[Si1]{$\Si$}
\obj(80,20)[LM1]{$L(M)$}
\mor{Si}{Si1}{$R_S$}
\mor{LM}{LM1}{$R_{i(S)}$}
\mor{Si}{LM}{$e$}
\mor{Si1}{LM1}{$e$}
\enddc
\fn$$
i.e.~$e \circ R_S = R_{i(S)} \circ e$, where $R$ denotes the relevant canonical right actions defined on the principal bundles $\Si$ and $L(M)$ 
and where $i:\SO(\eta)\arr \GL(m)$ is the canonical group inclusion.
We stress that on any $M$ which allows global metrics of signature $\eta$ the bundle $\Si$ can always be chosen so that there exist global frames; 
see \ref{Jadwisin}. 
Locally the frame is represented by invertible matrices $e_a^\mu$ and it defines a spacetime metric
$g_{\mu\nu}= e_a^\mu \>\eta_{ab}\> e_b^\nu$ which is called the {\it induced metric}.

As for the Levi-Civita connection, a frame defines a connection on $\Si$ (called the spin-connection of the frame) given by
$$
\om^{ab}_\mu= e^a_\al \( \Ga^\al_{\be\mu} e^{b\be} + d_\mu e^{b\al}\)
\fl{FrameConnection}$$
where $\Ga^\al_{\be\mu}$  denote Christoffel symbols of the induced metric.
The spin-connection is compatible with the frame in the sense that
$$
\na_\mu e^\nu_a = d_\mu e^\nu_a + \Ga^\nu_{\la\mu} e^\la_a - \om^c{}_{a\mu} e^\nu_c \equiv 0
\fn$$

In general the (natural) lift $\hat \xi$ of a spacetime vector field $\xi$ to $L(M)$ is not adapted to the image $e(\Si)\subset L(M)$ and thence 
it does not define any vector field on $\Si$.
%However, one can use shewsymmetrization of tetrad indices to define an adapted vector field.
With this notation the Kosmann lift of $\xi=\xi^\mu\del_\mu$  is defined by $\hat\xi_K= \xi^\mu\del_\mu + \hat \xi^{ab}\si_{ab}$ (see \ref{Godina}) where we set:
$$
\hat \xi^{ab}=  e_\nu^{[a} \na_\mu\xi^\nu   e^{b]\mu} -\om^{ab}_\mu \xi^\mu
\fl{KosmannLift}$$ 
and where $e^{a\mu}=\eta^{ac}e_c^\mu$ and $e_\nu^b$ denote the inverse frame matrix.

Let us stress that despite appearing so, the Kosmann lift \ShowLabel{KosmannLift} does not in fact depend on the connection, but just on the frame and its first derivatives. The same lift can be written as $\hat \xi^{ab}=  \na^{[b}\xi^{a]}   -\om^{ab}_\mu \xi^\mu$ where we set $\xi^a=\xi^\mu e^a_\mu$
since one can prove that
$$
\na_b \xi^a=e_\nu^{a} \na_\mu\xi^\nu   e^{\mu}_b
\fn$$

Another useful equivalent expression for the Kosmann lift is giving the vertical part of the lift with respect to the spin connection (see \ref{Libro}, pages 288--290), namely
$$
\hat \xi^{ab}_{(V)}:= \hat \xi^{ab} +  \om^{ab}_\mu \xi^\mu =  e_\nu^{[a} \na_\mu\xi^\nu   e^{b]\mu}= \na^{[b} \xi^{a]}
\fl{KLV}$$
This last expression is useful since it expresses a manifestly covariant quantity.

We have to stress that the Kosmann lift does not preserve commutators. In fact if one  considers two spacetime vectors $\xi$ and $\ze$ and 
computes the Kosmann lift of the commutator $[\xi, \ze]$ one can easily prove that
$$
[\xi, \ze]\>{}\hat{}_{K}= [\hat \xi_K, \hat \ze_K] + \frac[1/2] e^a_\al \Lie_\ze g^{\al\la} \Lie_\xi g_{\la\be} e^{b\be} \si_{ab}
\fn$$
Thence only if one restricts to Killing vectors (i.e.~$\Lie_\xi g=0$) one recovers that the lift preserves commutators.

\NewSection{The Lie Derivative of Lorentz Tensors}

Let $\la$ be a representation (of rank $(p,q)$) of $\SO(\eta)$ over a suitable vector space $V$. 
Let $E_A$ be a basis of $V$ so that a point $t\in V$ is given by $t=t^A E_A$ and $\la(J,t)= \la^A_B(J) t^B$.

\Note
For example, if $V=T^1_1(\R^m)\sim \R^m\otimes\R^m$ with coordinates $t^a_b$ we may have 
$$
\la:\SO(\eta)\times V\arr V:(J, t)\mapsto J^a_c t^c_d \bar J^d_b
\fn$$
the bar denoting now the inverse in $\SO(\eta)$. This is the tensor representation of rank $(1,1)$.
\endNote

Then, by definition, a {\it Lorentz tensor} is a section of the  bundle  $\Si_\la=\Si\times_\la V$ 
associated to $\la$ through the representation $\la$.
Fibered coordinates on $\Si_\la$ are in the form $(x^\mu, t^A)$ and transition functions of $\Si$ act on $\Si_\la$ through the representation $\la$.

If we consider a global infinitesimal  generator of automorphisms over $\Si$ (also called a {\it Lorentz transformation}) locally expressed as
$$
\Xi= \xi^\mu(x)\del_\mu + \xi^{ab}(x)\si_{ab}
\fn$$
(which projects over the spacetime vector field $\xi=\xi^\mu\del_\mu$) 
this induces a global vector field over $\Si_\la$ locally given by
$$
\Xi_\la=\xi^\mu(x)\del_\mu + \xi^{A}{\del\over \del t_A}
\qquad
\xi^A= \xi^{ab} \del_{ab}\la^A_B(\one) t^B
\fn$$
Let us remark that this vector field is linear in $\xi$.

\Note
For example,  if $\la$ is the tensor representation of rank $(1,1)$ given above, then the induced vector field
is 
$$
\Xi_\la=\xi^\mu \del_\mu + \(\xi^{a}\_c \>t^c_b- t^a_d\> \xi^d\_b\){\del\over \del t^a_b}
\fn$$
where indices are lowered and raised by $\eta_{ab}$.
\endNote

According to the general framework for Lie derivatives (see \ref{Trautman}) for a section $t: M\arr \Si_\la:x^\mu\mapsto(x,  t^A(x))$ of the bundle
$\Si_\la$ with respect to the (infinitesimal) Lorentz tranformation $\Xi$, we find
$$
\Lie_\Xi t= Tt(\xi)-\Xi_\la\circ t 
= \(\xi^\mu d_\mu t^A- \xi^{ab} \del_{ab}\la^A_B(\one) t^B\) {\del\over \del t^A} 
\fl{gaugeLieDerivative}$$

\Note
For example,  if $\la$ is the tensor representation of rank $(1,1)$ given above the Lie derivative of a section reads as
$$
\eqalign{
\Lie_\Xi t=&\(\xi^\mu d_\mu t^a_b- \xi^{a}\_c \>t^c_b+ t^a_d\> \xi^d\_b\) {\del\over \del t^a_b}=
%\cr=&
\(\xi^\mu \na_\mu t^a_b- (\xi_{(V)})^{a}\_c \>t^c_b+ t^a_d\> (\xi_{(V)})^d\_b\) {\del\over \del t^a_b}\cr
}
\fn$$
where $ (\xi_{(V)})^{a}\_c=  \xi^{a}\_c + \om^{a}{}_{c\mu} \xi^\mu$ denotes the vertical part of $\Xi$ with respect to the same connection used for the covariant derivative $\na_\mu t^a_b= d_\mu t^a_b + \om^{a}{}_{c\mu} t^c_b - \om^c{}_{b\mu} t^a_c $.
Let us stress that in spite of its convenient connection-dependent expressions the Lie derivative does not eventually depend on any connection (as it may seem from our second expression).
\endNote

Notice that this definition of Lie derivatives is natural, i.e.~it preserves commutators, namely
$$
[\Lie_{\Xi_1}, \Lie_{\Xi_2}]\si = \Lie_{[\Xi_1, \Xi_2]}\si
\fl{naturality}$$
Unfortunately, Lorentz tranformations as introduced above have nothing to do with coordinate transformations (or spacetime diffeomorphisms).
They have been introduced as gauge transformations acting pointwise and completely unrelated to spacetime diffeomorphisms.
Indeed the Lie derivative \ShowLabel{gaugeLieDerivative} can be performed with respect to bundle vector fields $\Xi$ instead of spacetime
vector fields and this is completely counterintuitive if compared with what expected for spacetime objects like, for example, spinors.
These objects are in fact expected to react to spacetime transformations; on the other hand, on a general spacetime there is nothing like Lorentz transformations.

We shall hence define Lie derivatives of Lorentz tensors with respect to any spacetime vector field and then show that in Minkowski spacetime, where Lorentz trasformations are defined, these reproduce and extend the standard notion.
The price to be paid is loosing naturality like \ShowLabel{naturality} (which will be retained only for Killing vectors if Killing vectors exist on $M$).

Let us restrict to vector fields $\hat\xi_K$ of $\Si$ which are the Kosmann lift of a spacetime vector field
$\xi$ and define the Lie derivative of the  Lorentz tensor $t$ with respect to the spacetime vector field $\xi$
to be
$$
\Lie_\xi t\equiv \Lie_{\hat \xi_K} t = (\xi^\mu d_\mu t^A- \hat \xi^{ab} \del_{ab}\la^A_B(\one) t^B) {\del\over \del t^A} 
\fn$$ 
where $\hat \xi^{ab}$ is expressed in terms of the derivatives of $\xi^\mu$ (and the frame) as in \ShowLabel{KosmannLift}.

\Note
For example, for Lorentz tensors of rank $(1,1)$ we have
$$
\eqalign{
\Lie_\xi t\equiv \Lie_{\hat \xi} t=& \(\xi^\mu d_\mu t^a_b- \hat \xi^{a}\_c \>t^c_b+ t^a_d\> \hat \xi^d\_b\) {\del\over \del t^a_b}
=\(\xi^\mu \na_\mu t^a_b- (\hat \xi_{(V)})^{a}\_c \>t^c_b+ t^a_d\> (\hat \xi_{(V)})^d\_b\) {\del\over \del t^a_b}=\cr
=& \(\xi^\mu \na_\mu t^a_b- \na_c\xi^a\>t^c_b+ t^a_d\> \na_b\xi^d\) {\del\over \del t^a_b}
=\(  \na_d \(\xi^dt^a_b\) - \na_c\xi^a\>t^c_b\) {\del\over \del t^a_b}
\cr
}
\fn$$ 
For a generic Lorentz tensor of any rank, similar terms arise one for each Lorentz index. 
\endNote

Now since the Kosmann lift on $\Si$ does not preserve commutators these Lie derivatives are not natural unless
one artificially restricts $\xi$ to be a Killing vector (of course provided $M$ allows Killing vectors!). 
In fact, one has generically
$$
\Lie_{[\xi, \ze]} t\equiv  \Lie_{[\xi, \ze]\>\hat{}_K} t \not=   \Lie_{[\hat\xi_K, \hat\ze_K]} t= [\Lie_{\hat \xi_K}, \Lie_{\hat \ze_K}] t\equiv [\Lie_\xi, \Lie_\ze] t
\fn$$

\Note
One can try to specialize this to simple cases in order to make non--naturality manifest.
For example, if one considers a Lorentz vector $v^a$  and two spacetime vector fields $\xi$ and $\ze$ one can easily check that
$$
\Lie_{[\xi, \ze]} v^a =[\Lie_\xi, \Lie_\ze] v^a 
+\frac[1/4] \(v^\al g^{\be\rho} e^{a\si} - v^\rho g^{\be\si} e^{a\al}\) \Lie_\xi g_{\rho\si} \Lie_\ze g_{\al\be}
\fn$$ 
Let us remark that according to this expression when $\xi$ or $\ze$ are Killing vectors of the metric $g$ commutators are preserved.
Moreover, the extra term does not vanish in general.

Of course, there are degenerate cases (e.g.~setting $\xi=\ze$) in which the extra terms vanishes due to coefficients without requiring Killing vectors.
However, in this case also the other terms vanish.
\endNote

\NewSection{Properties of Lie Derivatives of Lorentz Tensors }

We shall prove here two important properties of Lie derivatives as defined above
(see, for example, \ref{Vandyck2}, \ref{Vandyck3}, \ref{Mtheo}, \ref{Vandyck1} and references quoted therein) 

For the Lie derivative of a frame one has
$$
\Lie_\xi e^a_\mu= \xi^\la \na_\la e^a_\mu -\na_\mu\xi^\la e^a_\la + (\hat\xi_{(V)})^a_b e^b_\mu
\fn$$
If we are using, as we can always choose to do, the spin and the Levi-Civita connections for the relevant covariant derivatives, then $\na_\la e^a_\mu=0$.
By using the Kosmann lift \ShowLabel{KLV} one easily obtains
$$
\eqalign{
\Lie_\xi e^a_\mu=& -\na_\mu\xi^\la e^a_\la +\na^{[b}\xi^{a]} e_{b\mu}=
-\na_\mu\xi^\la e^a_\la +\na_{[\mu} \xi_{\la]} e^{a\la} = 
-\na_{(\mu} \xi_{\la)} e^{a\la}=\cr
=&\frac[1/2] \Lie_\xi g_{\mu\la} e^{a\la}\cr
}
\fn$$
This expression holds true for any spacetime vector $\xi$ and of course it proves that the Lie derivative  vanishes along Killing vectors.

Let us stress that this last expression, obtained here from the general prescription for the Lie derivative of Lorentz tensors, is trivial in view of the expression on the induced metric as a function of the frame; in fact,
$$
\frac[1/2] \Lie_\xi g_{\mu\la} e^{a\la}= \Lie_\xi e^c_\mu e_{c\la}e^{a\la}=\Lie_\xi  e^a_\mu 
\fn$$

For the second property we wish to prove let us first notice that the frame induces an isomorphism between $TM$ (on which one considers $(x^\mu, v^\mu)$ as fibered coordinates) and the bundle of Lorentz vectors
$\Si\times_\la \R^m$ (on which $(x^\mu, v^a)$ are considered as  fibered coordinates) by
$$
\Phi:TM \arr \Si\times_\la \R^m: v^\mu\mapsto v^a=e^a_\mu v^\mu 
\fn$$
We can thence express the Lie derivative of a section $v$ of $\Si\times_\la \R^m$ (i.e.~a Lorentz vector) in terms of the Lie derivative of the corresponding section of $TM$.
In fact one has:
$$
\eqalign{
\Lie_\xi v^a=& \xi^\mu \na_\mu v^a -(\hat \xi_{(V)})^a_b v^b=
\xi^b \na_b v^a -\na^{[b}\xi^{a]} v_b=\Lie_\xi v^\mu e_\mu^a +\na^{(b}\xi^{a)} v_b=\cr
=&\Lie_\xi v^\mu e_\mu^a -\frac[1/2]e^a_\mu \Lie_\xi g^{\mu\nu}e^b_\nu v_b=
\Lie_\xi v^\mu e_\mu^a +\Lie_\xi e^a_\mu  v^\mu
\cr
}
\fl{VectorLieDerivative}$$
Let us stress that these two properties hold true for any spacetime vector field $\xi$ and they specialize to the ones discussed in \ref{Ortin} for Killing vectors.

The origin and meaning of the Lie derivative \ShowLabel{VectorLieDerivative} can be easily understood: one has to take into account that if one drags $\xi^a$ along a vector field
the overall change of the object receives a contribution from how the vector changes but also a contribution from how the frame changes.

Similar properties can be easily  found for Lorentz tensors of any rank since the frame transforms ordinary tensors into Lorentz tensors; e.g.~one has
$$
\Phi: t^\mu_\nu\mapsto t^a_b = e^a_\al t^\al_\be e^\be_b
\fn$$

\NewSection{Transformation of Lorentz Vectors in Minkowski Spacetime}

Let us consider Minkowski spacetime $M=\R^4$ with the metric $\eta$; being it contractible any bundle over it is trivial. As a consequence  we are forced to choose $\Si=\R^4\times \SO(3,1)$. Since $M\equiv \R^4$ is parallelizable, its frame bundle  is trivial, i.e.~$L(\R^4)=\R^4\times \GL(4)$. 
Let us fix Cartesian coordinates $x^\mu$ on $M\equiv \R^4$ and let us fix a frame $e_a= \de_a^\mu \del_\mu$; such a  frame  induces the Minkowski metric $\eta_{\mu\nu}$.

In such notation the Levi-Civita connection vanishes, $\Ga^\al_{\be\mu}=0$ and the spin connection too,  $\om^{ab}_\mu=0$; the Kosman lift hence specializes to
$$
\hat\xi^{ab}_{(V)}= e^{[b\be}\na_\be\xi^\al e^{a]}_\al
\fn$$

Let us now consider a vector field $\xi$ the flow which is made of Lorentz coordinate tranformations $x'^\mu=\La^\mu_\nu x^\nu$; 
since $\xi$ is of course a Killing vector, then the Lie derivative of a Lorentz vector is
$$
\Lie_\xi v^a=\Lie_\xi v^\mu e_\mu^a =\( \xi^\al \del_\al v^\mu - v^\al \dot\La^\mu_\al \) \de_\mu^a
\fl{52}$$
Such a Lie derivative corresponds to the trasformation rules
$$
v'^a=\La^a_b v^b
\fl{53}$$
which is exactly as a vector is expected to trasform under a Lorentz coordinate transformation.

A similar result can be easily extended to covectors, tensors and, with slight though obvious changes, to spinors.
When $\xi$ is not Killing, however, the Lie derivative may not be the infinitesimal counterpart of a finite transformation rule as in \ShowLabel{52} and \ShowLabel{53}; 
in this case the traditional interpretation of Lie derivatives as a measure of changing of objects dragged along spacetime vector fields
fails to hold true. One should however wonder whether such an interpretation is really fundamental to many common uses of Lie derivatives.
Our answer is in the negative as one can argue by a detailed analysis of physical quantities containing Lie derivatives.

Lie derivatives appear, e.g., in Noether theorem; 
in this case they appear naturally as a by--product of variational techniques.
Here Noether currents turn out to be expressed in terms of Lie derivatives expressed as in equation \ShowLabel{gaugeLieDerivative}.
The interpretation of such Lie derivatives as measuring infinitesimal changes along symmetry transformations is important since, based on that, one can relate
Noether currents to symmetries.

Now the essential point is that there is no reason to expect spacetime vector fields to be the most general (infinitesimal) symmetries in Physics.
Fundamentally speaking, symmetries encode the observers' freedom to set their conventions to describe Physical world. 
While coordinates are certainly necessary conventions for any observer (and hence general covariance principle is a fundamental symmetry that should be expected in any physical system), special systems might need further conventions which might result in independent class of symmetries (as it happens in gauge theories, e.g.~electromagnetism).

Of course, since these further conventions are independent of spacetime coordinate fixing, gauge transformations cannot be expressed as spacetime diffeomorphisms, but they are expressed as field transformations. As such they are vector fields on the configuration bundle, not on spacetime. It is hence reasonable and important to have a notion of Lie derivative of fields along bundle vectors, as in \ShowLabel{gaugeLieDerivative}. 
It is only in GR where symmetries come from spacetime vector fields that one should expect Lie derivatives along spacetime vector fields and their interpretation as quantities related to the spacetime geometry. 

This more general situation, i.e.~when the quantities entering Noether theorem are interpreted as Lie derivatives of fields along bundle vectors, can be 
simply discussed by considering a very well-known physical situation, i.e.~covariant electromagnetism.

The electromagnetic field $F_{\mu\nu}= \del_\mu A_\nu-  \del_\nu A_\mu$ is the curvature of a field $A_\mu$ which is usually known as a {\it quadripotential}
and, as it is well known, is a connection on a principal bundle $P$ for the group $U(1)$. This is the standard gauge approach to electromagnetism. 
The Maxwell Lagrangian is
$$
L_{M}= -\frac[1/4] \sqrt{g} F_{\mu\nu} F\^\mu\^\nu
\fl{EMLag}$$
By variation one obtains
$$
\de L_{M}=  -\frac[1/2] \sqrt{g} H_{\al\be} \de g^{\al\be}  + \na_\mu \( \sqrt{g} F^{\mu\nu}\) \de A_\nu- \na_\mu \( \sqrt{g} F^{\mu\nu} \de A_\nu\)
\fl{EMVar}$$
where we set $H_{\al\be} = F_{\mu\al} F\^\mu{}_{\be} - \frac[1/4]  F_{\mu\nu} F\^\mu\^\nu g_{\mu\nu}$ for the standard energy-momentum tensor of the electromagnetic field.
The second term in \ShowLabel{EMVar} produces Maxwell equations, namely $\na_\mu \( \sqrt{g} F^{\mu\nu}\)=0$.
The third term relates to conservation laws (see \ref{Libro}).

The Lagrangian \ShowLabel{EMLag}  is covariant with respect to the infinitesimal transformations
$$
\Xi= \xi^\mu \Frac[\del/\del x^\mu] +2 \del_\al \xi^\mu g^{\al\nu} \Frac[\del/\del g^{\mu\nu}]
+ \( \del_\mu\xi-\del_\mu \xi^\nu A_\nu\) \Frac[\del/\del A_\mu] 
\fn$$
which correspond to  $1$-parameter families of gauge transformations
$$
\cases{
&x'^\mu=x'^\mu_{(\ep)}(x)\cr
&g'^{\mu\nu}=  \Frac[\del x'^\mu_{(\ep)}/\del x'^\al] g^{\al\be}  \Frac[\del x'^\nu_{(\ep)}/\del x'^\be]  \cr
&A'_\mu= \Frac[\del x^\nu/\del x'^\mu_{(\ep)}]\( A_\nu + \del_\nu \al_{(\ep)}\)\cr
}
\fn$$
Here the generator $\xi^\mu$ is related to the coordinate change $x'^\mu=x'^\mu_{(\ep)}(x)$ while the generator $\xi$ is related to the gauge transformation $\al_{(\ep)}$.

Let us remark that $\Xi$ is a vector field on the configuration bundle (that is a manifold with coordinates $(x^\mu, g^{\mu\nu}, A_\mu)$), not on spacetime.
In a general situation (namely unless the principal bundle $P$ is assumed to be trivial) there is no way of either lifting a spacetime vector field
to the configuration bundle or globally setting $\xi=0$ so to split the vector $\Xi$
into a spacetime vector and a {\it ``gauge generator''}. In a physical language one usually says that the condition $\xi=0$ is not gauge covariant and hence local, unless there exist global gauges. (By the way, also when global gauges exist, the condition is not gauge covariant and hence unphysical, from a fundamental viewpoint.)

The Lie derivative of the field $A_\mu$ along the symmetry generator $\Xi$ is in this case (see \ShowLabel{gaugeLieDerivative})
$$
\Lie_\Xi A_\mu= \xi^\la F_{\la\mu} - \na_\mu \( \xi-\xi^\la A_\la\)
\fn$$

Noether theorem in this case shows (see again \ref{Libro}) on-shell conservation of the following Noether current
$$
\eqalign{
\calE^\mu=&-\sqrt{g} \(F^{\mu\nu} \Lie_\Xi A_\nu +\xi^\mu L_{M} \)\cr
}
\fn$$
In the special case when $\xi^\mu=0$ one has
$$
\eqalign{
\calE^\mu=&\sqrt{g} \(F^{\mu\nu}  \na_\mu \xi  \)=  \na_\mu \(\sqrt{g}  F^{\mu\nu} \xi  \)- \na_\mu \(\sqrt{g}  F^{\mu\nu} \) \xi \cr
}
\fn$$
The second term vanishes on-shell, thus one obtains
$$
\calE^\mu= \na_\mu \(\sqrt{g}  F^{\mu\nu} \xi  \)
\fn$$
The corresponding conserved quantity is
$$
Q (\xi)= \Frac[1/2]\int_{\del \Om}  \sqrt{g}  F^{\mu\nu} \xi  \>ds_{\mu\nu}
\fn$$
where $ds_{\mu\nu}$ is the area element on the boundary of the $3$-region $\Om$ of spacetime. This is the electric charge defined {\it \`a la} Gauss.

This example shows clearly what happens in general when gauge transformations are allowed and symmetry generators live at bundle level:
also in this case Noether theorem involves Lie derivatives, though in the generalized sense introduced above.
In this case we are not dealing with Lorentz objects so one cannot introduce Kosmann lift (or similar lifts) and reduce everything to spacetime vector fields.

\NewSection{Applications}

In order to provide an example of concrete aplication of our formalism here introduced in action we shall here consider the application to the so called {\it Holst's action principle} (see \ref{Holst}) 
which is used as an equivalent formulation of GR suitable for developing LQG through the use of  the Barbero-Immirzi connection (see \ref{Barbero}, \ref{Immirzi},  \ref{myBI},  \ref{Rov2} as well as references quoted therein).

Let us first consider tetrad-affine formulation of GR: the fundamental fields are a Lorentz connection $\Ga^{ab}_\mu$ and a vielbein $e^a=e^a_\mu\>dx^\mu$.
The connection defines the curvature form $R^{ab}= \frac[1/2] R^{ab}{}_{\mu\nu} \>dx^\mu\land dx^\nu$.
Let us also set $e=\det |e^a_\mu|$, $R^a{}_\mu= R^{ab}{}_{\mu\nu} e_b^\nu$ and $R= R^{ab}{}_{\mu\nu} e_a^\mu e_b^\nu$;
here $e_b^\nu$ denotes the inverse frame matrix of  $e^b_\nu$.
The frame also defines a metric $g_{\mu\nu}=  e^a_\mu \eta_{ab}  e^b_\nu$ which in turn defines its Levi-Civita spacetime connection $\Ga^\al_{\be\mu}$.

On a spacetime of dimension $4$, let us consider the Lagrangian
$$
L_{tA}= R^{ab}\land e^c \land e^d\>\ep_{abcd}
\fn$$
By variation we obtain
$$
\de L_{tA}= -2e e^\si_a\( R^a{}_\mu -\frac[1/2] R e^a_\mu \) e_d^\mu \>\de e^d_\mu
- \ep_{abcd}\na_\mu\(e^c_\rho e^d_\si\)\ep^{\mu\nu\rho\si} \de\Ga^{ab}_\mu
+ \ep_{abcd}\na_\mu\(e^c_\rho e^d_\si \de\Ga^{ab}_\mu\)\ep^{\mu\nu\rho\si}
\fn$$
 
Thus one obtains field equations
$$
\cases{
&R^a{}_\mu -\frac[1/2] R e^a_\mu=0\cr
&\na_{[\mu}\(e^{[c}_\rho e^{d]}_{\si]}\)=0\cr
}\fn$$
The second field equation forces the connection to be the connection induced by the frame $\Ga^{ab}_\mu=\om^{ab}_\mu$ (see eq. \ShowLabel{FrameConnection}); then the first equation force the induced metric to obey Einstein equations.

This field theory is dynamically equivalent to standard GR, in the sense that it obeys equivalent field equations.
However, the theory is in fact richer in its physical interpretation, since the use of different variables and action principles generate larger symmetry and extra conservation laws.
In fact, this theory has a bigger symmetry group being generally covariant and Lorentz covariant.

Noether theorem impies then conservation of the current 
$$
\calE^\mu= 4e e_a^\mu e_b^\nu \Lie_\Xi \Ga^{ab}_\nu -\xi^\mu L_{tA}
\fn$$
along any Lorentz gauge generator $\Xi= \xi^\mu\del_\mu + \xi^{ab} \si_{ab}$.
The Lie derivative of a connection is given by
$$
\Lie_\Xi \Ga^{ab}_\nu= \xi^\la R^{ab}{}_{\la\nu} + \na_\nu \hat\xi^{ab}
\fn$$
where we set $\hat\xi^{ab}=\xi^{ab}+\xi^\la\Ga^{ab}_\la$.
 
Hence one obtains
$$
\eqalign{
\calE^\mu=&  4e e_a^\mu \(R^a{}_\mu-\frac[1/2]R e^a_\mu\) \xi^\la -  4\na_\nu\(e e_a^\mu e_b^\nu\)\hat\xi^{ab} + 4 \na_\nu\(e e_a^\mu e_b^\nu\hat\xi^{ab} \) 
}
\fn$$
The first and second terms vanish on-shell; hence one obtains
$$
\calE^\mu= 4 \na_\nu\(e e_a^\mu e_b^\nu\hat\xi^{ab} \) 
\fl{FreeNoetherCurrent}$$
Let us stress that this current depends only on the Lorentz generator $\hat\xi^{ab}$.

Here is the issue with physical interpretation: we have two equivalent formulations of Einstein  GR and Noether currents in one case depend on spacetime vector fields
while in tetrad--affine formulation  Noether currents depend on Lorentz generator which {\it a priori} has nothing to do with spacetime transformations.
Let us stress of course that unless the spacetime is Minkowski, there is no class of spacetime diffeomorphisms representing {\it Lorentz transformations}.

Considering the dynamical equivalence at level of field equations and solution space, one would like this equivalence to be extended 
at level of conservation laws.
Moreover, some of the conserved quantities in standard GR are known to be related to physical quantities such as energy, momentum and angular momentum,
while one would wish to be able to identify the corresponding quantities in the second formulation.
Kosmann lift is in fact essential to relate Lorentz generators to spacetime  diffeomorphisms and the corresponding conservation laws.

The Noether current \ShowLabel{FreeNoetherCurrent} can be restricted setting $\Xi=\hat \xi_K$ so that one obtains
$$
\calE^\mu_{tA}= 4 \na_\nu\(e \na^\mu \xi^\nu  \) 
\fn$$
which corresponds to the standard conserved quantity associated to spacetime diffeomorphisms in GR written in terms of Komar superpotential.
This (and only this) restores the equivalence between standard GR and tetrad--affine  formulation at level of conservation laws.

As a further example let us consider the covariant Lagrangian:
$$
L_H= L_{tA} + \be R^{ab}\land e_a\land e_b 
\fn$$
which is known as Holst's Lagrangian.

By variationas one obtains equations
$$
\cases{
&e^\mu_d\(R^a_\mu -\frac[1/2] R e^a_\mu \) e_a^\si- \be R_{d\rho\mu\nu} \ep^{\mu\nu\rho\si}=0\cr
&\na_{[\mu}\(e^{[c}_\rho e^{d]}_{\si]}\)=0\cr
}
\fn$$
The second equation still imposes $\Ga^{ab}_\mu=\om^{ab}_\mu$; this in turns implies $R^a{}_{[\rho\mu\nu]}=0$ (first Bianchi identity)
and hence Einstein equations. This shows how also Holst's Lagrangian provides an equivalent formulation of standard GR.

It is interesting to check if also in this case the equivalence is preserved also at level of conservation laws.
The Noether current is
$$
\calE_{H}^\mu= 4e e_a^\mu e_b^\nu \Lie_\Xi \Ga^{ab}_\nu  + e e_c^\mu e_d^\nu \ep^{cd}\_a\_b \Lie_\Xi \Ga^{ab}_\nu- \xi^\mu L_{H}
\fn$$ 
As in the previous case this can be recasted modulo terms vanishing on-shell as follows
$$
\calE_{H}^\mu- \calE_{tA}^\mu = \na_\nu\(e e_c^\mu e_d^\nu \ep^{cd}\_a\_b \hat \xi^{ab}\)
\fn$$
Again this has nothing to do with sacetimes symmetries and in general would affect conserved quantities.
When Kosmann lift is again inserted into these conservation laws one obtains
$$
 \calE_{H}^\mu- \calE_{tA}^\mu = \na_\nu\( \na^\rho\xi^\si \ep^{\mu\nu}{}_{\rho\si} \)
\fn$$
which vanishes being the divergence of a divergence.
Hence once again the correspondence at level of conservation laws is preserved when the Kosmann lift is used.

\NewSection{Conclusion}

We presented a framework to deal with Lorentz objects and showed how it applies to tetrad--affine formulation and Holst's formulation of GR.
In particular we showed that equivalence can be extended at the level of conservation laws if one introduces the Kosmann lift which establishes a correspondence among symmetry generators in different formulations.

One could argue whether the Lie derivatives defined above could be physically interpreted in a correct way.
Of course, one could always restrict to situations in which enough Killing vectors exist (or even to Minkowski spacetime $(\R^4,\eta)$); 
in these cases the standard results are obtained in particular.

However, in a generic spacetime $(M, g)$ one has no Killing vectors and at the end one has to decide whether a physical interpretation 
of these objects along generic spacetime vector field makes any sense.

The framerwork we introduced for Lorentz tensors provides a rigorous way of investigating formal properties
which in our opinion are the only necessary basis for a physical intepretation of Lie derivatives of Lorentz tensors themselves.

%%%%%%%%%%%%%%%%%%%%%%%%%%%%%%%%%%%%%%%%%%%%%%%%%%%
\Acknowledgements

This work is partially supported by MIUR: PRIN 2005 on {\it Leggi di conservazione e termodinamica in meccanica dei continui e teorie di campo}.  
We also acknowledge the contribution of INFN (Iniziativa Specifica NA12) and the local research funds of Dipartimento di Matematica of Torino University.

\ShowBiblio
\end